\documentclass[twocolumn,prb,citeautoscript,showpacs]{revtex4-1}

\usepackage{amsmath}
\usepackage{amssymb}
\usepackage{graphicx}
\usepackage{hyperref}
\usepackage{soul}

\begin{document}

\title{Posner molecules: From atomic structure to nuclear spins}

\author{Michael W. Swift}
\affiliation{Department of Physics, University of California, Santa Barbara, California 93106-9530, USA}

\author{Chris G. Van de Walle}
\affiliation{Materials Department, University of California, Santa Barbara, California 93106-5050, USA}

\author{Matthew P. A. Fisher}
\affiliation{Department of Physics, University of California, Santa Barbara, California 93106-9530, USA}
\email{}

\newcommand{\Pos}{\ensuremath{\text{Ca}_9(\text{PO}_4)_6}}
\newcommand{\dbrac}[2]{\left\langle\left\langle #1 ; #2 \right\rangle\right\rangle}
\newcommand{\ket}[1]{\left| #1 \right\rangle}
\newcommand{\bra}[1]{\left\langle #1 \right|}
\newcommand{\vsigma}{\boldsymbol{\sigma}}

\begin{abstract}

We investigate ``Posner molecules'', calcium phosphate clusters with chemical formula \Pos.  Originally identified in hydroxyapatite, Posner molecules have also been observed as free-floating molecules \emph{in vitro}.  The formation and aggregation of Posner molecules have important implications for bone growth, and may also play a role in other biological processes such as the modulation of calcium and phosphate ion concentrations within the mitochondrial matrix.  In this work, we use a first-principles computational methodology to study
the structure of Posner molecules, their vibrational spectra, their interactions with other cations, and the process of pairwise bonding.  Additionally, we show that the Posner molecule provides an ideal environment for the
six constituent $^{31}\text{P}$ nuclear spins to obtain very long spin coherence times.  \emph{In vitro}, the spins could provide a platform for liquid-state nuclear magnetic resonance quantum computation.  \emph{In vivo}, the spins may have medical imaging applications.  The spins have also been suggested as ``neural qubits'' in a proposed mechanism for quantum processing in the brain.

\end{abstract}

\maketitle

\section{Introduction}

In 1975 Betts and Posner, while examining the x-ray crystal structure of bone mineral---hydroxyapatite, $\text{Ca}_{10} (\text{PO}_4)_6 (\text{OH})_2$---noticed that within each unit cell there were
two calcium-phosphate ``structural clusters" with atomic constituents $\text{Ca}_{9} (\text{PO}_4)_6$.~\cite{Posner75}
It was subsequently argued that these so-called ``Posner clusters" played an important role
in the formation of amorphous calcium phosphate, which can be viewed as a ``glass"
of Posner clusters.

It was over 20 years later that Onuma and Ito, while performing intensity-enhanced dynamic light scattering on simulated body fluids, identified evidence for free-floating calcium phosphate clusters of size roughly $1 \text{ nm}$.~\cite{Onuma98}
They suggested that these clusters, apparently stable in solution for months or longer were, in fact, Posner clusters.  No spontaneous precipitation was observed even in a supersaturated solution, but it was suggested that these Posner clusters play a role in bone (hydroxyapatite) nucleation when presented with a seed crystal~\cite{Yin03,Du13}.
Additional evidence was provided in 2010
when cryogenic transmission electron microscopy experiments on bone nucleation in simulated body fluids imaged free-floating nanometer-sized molecules which coalesced near a seed, forming amorphous calcium phosphate
before undergoing a dramatic transition into the crystalline form of hydroxyapatite.~\cite{Dey10}
Further evidence for the structural integrity of these clusters in solution was found
in AFM images of bone growth on calcite surfaces.~\cite{Wang12}  Taken together, these remarkable experiments provide evidence that Posner clusters are stable in solution and,
as such, should perhaps be called ``Posner molecules"---the name we will henceforth adopt.

\begin{figure}
\includegraphics[width=0.8\columnwidth]{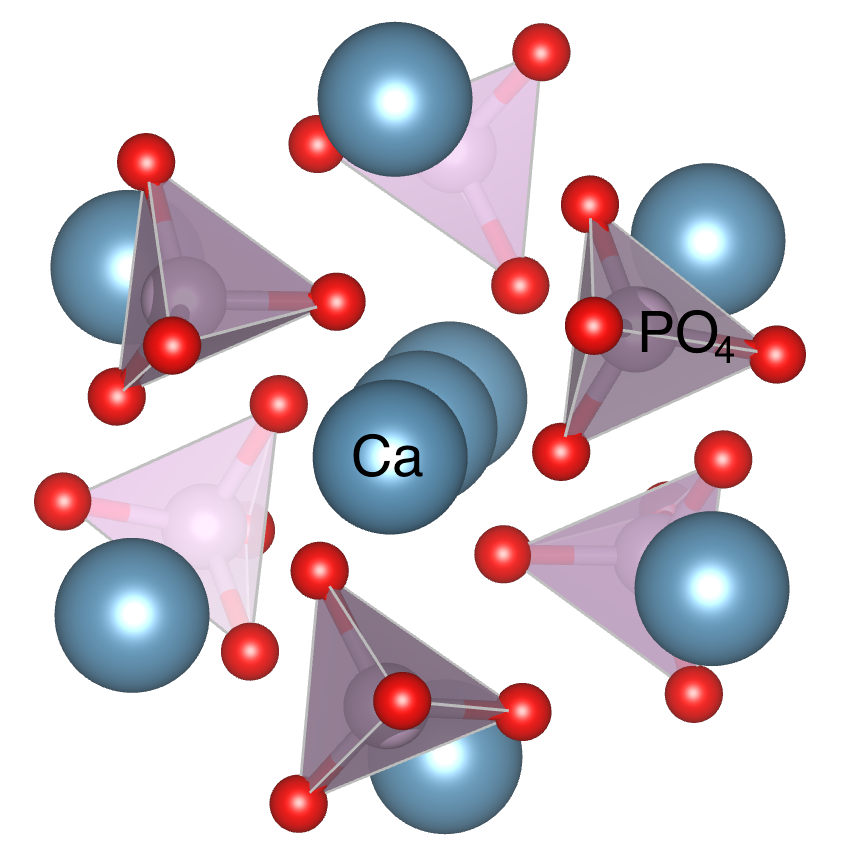}
\caption{Posner molecule, \Pos, with $\text{S}_6$ symmetry.  The symmetry axis is coming out of the page, tilted at a slight angle to show underlying calcium ions.  The symmetry axis may be thought of as the (111) axis of a cube.  A phosphate ion tetrahedrally coordinated by oxygens is at each face of the cube, a calcium ion is at each vertex, and a final calcium ion is in the center.  The cube is flattened by about 3\% along the symmetry axis.  Calcium ions are shown in blue, and phosphate ions are represented as purple tetrahedra with red oxygen ions at the vertices. }
\label{Posner}
\end{figure}

Soon after the Onuma and Ito experiments, several quantum chemistry calculations examined the putative arrangement of the ions in these Posner molecules, $\text{Ca}_{9} (\text{PO}_4)_6$, which were indeed found to be stable in vacuum.~\cite{Treboux00,Kanzaki01}
Their basic form consisted of eight calcium ions on the corners of a cube with the ninth calcium in the center, while the six phosphate ions were situated on the six faces of the cube.
Due to the symmetry incompatibility of the tetrahedral phosphate ions ($\text{PO}_4^{3-}$) with the cube faces,
the cubic symmetry was broken in the low-energy molecular configurations.
One of the most stable configurations was found to have $\text{S}_6$ symmetry, with a threefold rotational symmetry axis that is aligned along one of the cube diagonals, as depicted in Figure \ref{Posner}.

A more detailed exploration of Posner molecules is worthwhile for multiple reasons:

1. {\it Role in biological processes}.
Since experiments have provided strong evidence that Posner molecules are stable in simulated body fluids, it is likely that they are present
{\it in vivo}, particularly in the extracellular fluid where the free calcium concentration is appreciable.  These extracelluar Posner molecules could act as a reservoir for the previously suggested Posner-molecule-mediated mechanism for bone growth~\cite{Yin03,Du13}.
The molecules may also be present within cells.  Though cytoplasm is unlikely to contain Posner molecules due to its low calcium concentration, the mitochondiral matrix is known to contain stable calcium-phosphate complexes with a calcium:phosphate ratio suggestively close to that of Posner molecules (3:2).~\cite{Nicholls04}  Posner molecules may form a dynamic aggregate within the mitochondiral matrix,
with formation, aggregation, and dis-aggregation being finely modulated by the changing pH.~\cite{PrivateComm}

The Posner molecule is also central to the proposed ``quantum brain'' concept set forth in Ref.~\citenum{Fisher15}.  In this proposal, clouds of entangled Posner molecules in the brain act as ``neural qubits'', serving as a platform for quantum computation in cognitive processes.  Many of the properties we discuss here are key to the ongoing exploration of this concept.

2. {\it Long-lived phosphorus nuclear spin states}.
Molecules or ions with isolated nuclear spins that exhibit macroscopic coherence times are of great physical interest, both theoretically and practically.
Nuclear spins in liquid-state nuclear magnetic resonance (NMR) have some of the longest known coherence times of any system in physics,
especially nuclei with spin 1/2 that do not couple to electric fields and interact with the environment only through magnetic fields (e.g., dipole fields from other nuclear spins) and intra-molecular electron-mediated exchange interactions~\cite{NMR}.  Motional narrowing due to the rapid rotation of the molecules in solution averages out dipole-dipole magnetic field interactions, giving spin-1/2 nuclei long coherence times.~\cite{Mundy}

Since the most abundant isotopes of both calcium and oxygen have zero nuclear spin, we predict that
the six $^{31}\text{P}$ nuclear spins with S=1/2 in a Posner molecule are especially long lived, making them an intriguing platform for liquid-state NMR quantum computation~\cite{Vandersypen01}.
If the nuclear spins can be hyper-polarized, the spins could have medical imaging applications, since
the $^{31}\text{P}$ NMR signal is quite robust.   The Posner molecule's nuclear spins are also central to the quantum brain concept~\cite{Fisher15}, in which long coherence times are key to their role as neural qubits.

3. {\it Doping with other cations}.
We shall present evidence that replacing the calcium ion at the center of a Posner molecule with either another divalent cation or a pair of monovalent cations can further stabilize the Posner molecule.  This provides a possible mechanism for the known impact of elements such as lithium, magnesium, and iron on bone health~\cite{Zamani09,Toxqui15,Castiglioni13}.  Impurities are also relevant to the quantum brain hypothesis~\cite{Fisher15}, which suggests that the cognitive effects of lithium (and the isotope dependence of these effects) may arise from the interaction of the lithium nuclear spin with the neural qubits provided by the $^{31}\text{P}$ nuclear spins.

4. {\it Aggregation of Posner molecules.}
The presence of two Posner clusters within the unit cell of hydroxyapatite suggests the possible importance of aggregation of Posner molecules in bone growth.  Our quantum chemistry calculations provide evidence that a pair of Posner molecules can chemically bind together (in vacuum) with a substantial binding energy of roughly $5 \text{ eV}$.  This pairwise bonding is the first step towards larger-scale aggregation, which may lead to the formation of amorphous calcium phosphate and eventually hydroxyapatite.~\cite{Yin03,Du13,Dey10,Wang12}

5. {\it Quantum dynamics of the six phosphorus nuclear spins in a Posner molecule}.
Provided the Posner molecules do have an $\text{S}_6$  symmetry,
the $2^6 = 64$ nuclear spin eigenstates in each molecule can be chosen also as eigenstates
under a 3-fold rotation about the symmetry axis, with eigenvalues of the form
$e^{i 2\pi \tau/3}$ with the ``pseudospin" $\tau$ taking one of 3 allowed values, $\tau =0,\pm1$.  Understanding the dynamics of these spins is important for any application of Posner molecules to medical imaging or quantum computation as discussed above.  The spin dynamics are also key to the quantum brain concept~\cite{Fisher15}.

Our paper is organized as follows.  After a brief description of the computational methods (Sec.~\ref{methods}), we treat the structural properties of the Posner molecule: symmetry (Sec.~\ref{Symmetry}), vibrational spectra (Sec.~\ref{Vibration}), impurity incorporation (Sec.~\ref{impurities}), and pairwise bonding (Sec.~\ref{bonding}), discussing the results and implications of each in turn.  We then move on to the spin properties, using first-principles methods to calculate the indirect spin-spin coupling between $^{31}\text{P}$ nuclear spins in a Posner molecule.  These values become input to an effective Hamiltonian for the spins.  We study this effective model, paying special attention to the implications for coherence of encoded quantum information (Sec.~\ref{spin_coupling}).

\section{Computational Methods}{}
\label{methods}

Our first-principles calculations are based on density functional theory (DFT).

Molecular symmetry, impurity incorporation, and pairwise bonding were studied with the Projector-Augmented Wave (PAW) method~\cite{PAW1} as implemented in the Vienna Ab initio Simulation Package (VASP)~\cite{VASP4}.  The hybrid exchange-correlation functional of Heyd, Scuseria, and Ernzerhof~\cite{HSE03} was employed with the standard $25\%$ mixing and screening parameter $\omega = 0.20$ \AA$^{-1}$ (a combination of parameters commonly referred to as HSE06)~\cite{HSE06}.    These calculations used a plane-wave energy cutoff of 400 eV.
Single Posner molecule calculations were performed in a vacuum supercell 16 \AA\ on a side.  Calculations exploring the chemical bonding between two Posner molecules used a 16 {\AA}$\times$16 {\AA}$\times$32 {\AA} cell.

The vibrational spectrum of {\Pos} was calculated via density functional perturbation theory (DFPT) with the Quantum {\sc Espresso} package~\cite{Giannozzi09} using ultrasoft pseudopotentials~\cite{Vanderbilt90} and the Perdew-Burke-Ernzerhof exchange-correlation functional~\cite{Perdew96}.  The acoustic sum rule was applied for the diagonalization of the dynamical matrix in order to account for the molecule's translational and rotational degrees of freedom.

Calculations of the pairwise nuclear spin-spin interactions between $^{31}\text{P}$ nuclear spins inside a Posner molecule were performed in Dalton2015~\cite{Dalton1,Dalton2} with the B3LYP hybrid functional~\cite{Becke93} using the method of Refs.~\onlinecite{Helgaker00,Helgaker08}.  The electronic states around oxygen and calcium were expanded in the popular 6-31G** basis~\cite{Pople73,Pople98}, while those around phosphorus were expanded in the pcJ-4 basis, which is optimized for the calculation of indirect spin-spin couplings.~\cite{Jensen06}

Molecular visualizations were produced using VESTA.~\cite{VESTA}

\section{Results and Discussion}

\subsection{Symmetry}
\label{Symmetry}
Previous computational work~\cite{Treboux00,Kanzaki01} used DFT to study the possible symmetries of a Posner molecule.  The authors found various candidates for the lowest-energy molecular structure, all within the numerical accuracy of the technique.  They concluded that the $\text{S}_6$ structure, the candidate ground state with the highest symmetry, is the prototypical Posner molecule.

\begin{figure}
\includegraphics[width=\columnwidth]{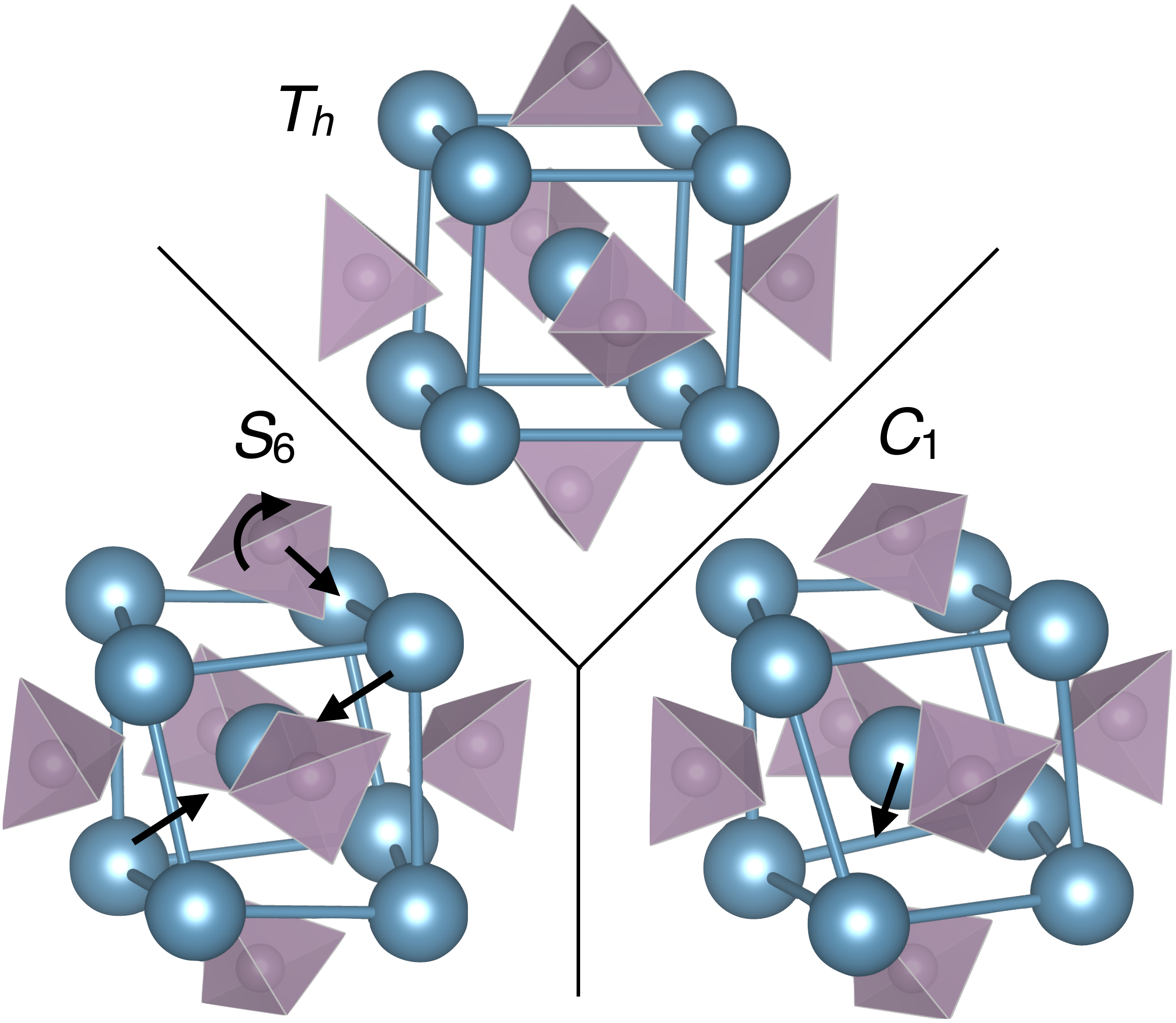}
\caption{Different symmetries of a Posner molecule, \Pos.  All structures were relaxed in vacuum.
The $\text{T}_h$ structure is unstable to distortion (compression along a diagonal of the cube and rotation of the phosphate tetrahedra), lowering the symmetry to $\text{S}_6$ about the compressed diagonal.  The reduction of symmetry from $\text{T}_h$ to $\text{S}_6$ lowers the energy substantially, by 1.98 eV.  While the $\text{S}_6$ structure is locally stable, a $\text{C}_1$ structure corresponding to an off-centering of the calcium atoms lowers the energy by only 0.06 eV.  As discussed in the text, we identify the $\text{S}_6$ structure as the prototypical Posner molecule. }
\label{fig:symmetry}
\end{figure}

Our calculations agree with these results.  Several structures are illustrated in Fig.~\ref{fig:symmetry}.  The lowest-energy structures have $\text{C}_1$ (no) symmetry but are only slightly distorted from $\text{S}_6$ symmetry.  The $\text{S}_6$ structure is higher in energy by only 0.06 eV, or less than 2 meV per atom, while the $\text{T}_h$ structure is a full 1.98 eV higher in energy than the $\text{S}_6$.  This is consistent with earlier work, which identified $\text{S}_6$ as the prototypical structure.~\cite{Kanzaki01}  To further test this identification, five $\text{C}_1$ structures were generated by random 0.1 \AA\ perturbations of an $\text{S}_6$ structure.  After relaxation, the atomic positions of these structures are found to deviate from $\text{S}_6$ symmetry by at most 0.008 \AA, while the average positions across all five structures deviate from $\text{S}_6$ symmetry by at most 0.002 \AA.  This demonstrates that the $\text{S}_6$ structure is indeed correct on average, and we expect thermal fluctuations will wash out any symmetry breaking.  We therefore assume the $\text{S}_6$ symmetry for the Posner molecule in the remainder of this work.

\subsection{Vibrational Spectra}
\label{Vibration}
Experiments find evidence for calcium phosphate clusters roughly 1 nm in size in simulated body fluid~\cite{Wang12,Grases14}.  While it is suspected that these clusters are Posner molecules~\cite{Yin03}, this has not been shown conclusively.  Spectroscopic probes such as infrared absorption spectroscopy, Raman scattering, or wide-angle X-ray scattering could identify the molecules definitively.  IR spectroscopy is based on the absorption of incoming photons resonant with a dipole-inducing vibrational mode of a molecule~\cite{Wilson}.  A calculation of vibrational modes and their associated dipole moments indicates the wavelengths of light that could be absorbed by the molecule.  The IR activity of the associated vibrational mode, which is proportional to the square of the induced dipole moment, corresponds to the strength of the peak in an absorption experiment.  Practically speaking, IR spectroscopy in solution is challenging, since the H$_2$O peak at 1575 cm$^{-1}$ with an IR activity of 1.659 (D/\AA)$^2$/amu~~\cite{Porezag96} tends to wash out the IR absorption signal of any species in aqueous solution.
Nevertheless, spectroscopic methods remain one of the best ways to conclusively identify Posner molecules, and calculations of vibrational spectra are an essential step in this process.

In addition to their spectroscopic relevance, the vibrational modes of the molecule also couple to the interactions between the six phosphorus nuclear spins, which will be discussed in section~\ref{spin_coupling}.

\begin{figure}
\includegraphics[width=\columnwidth]{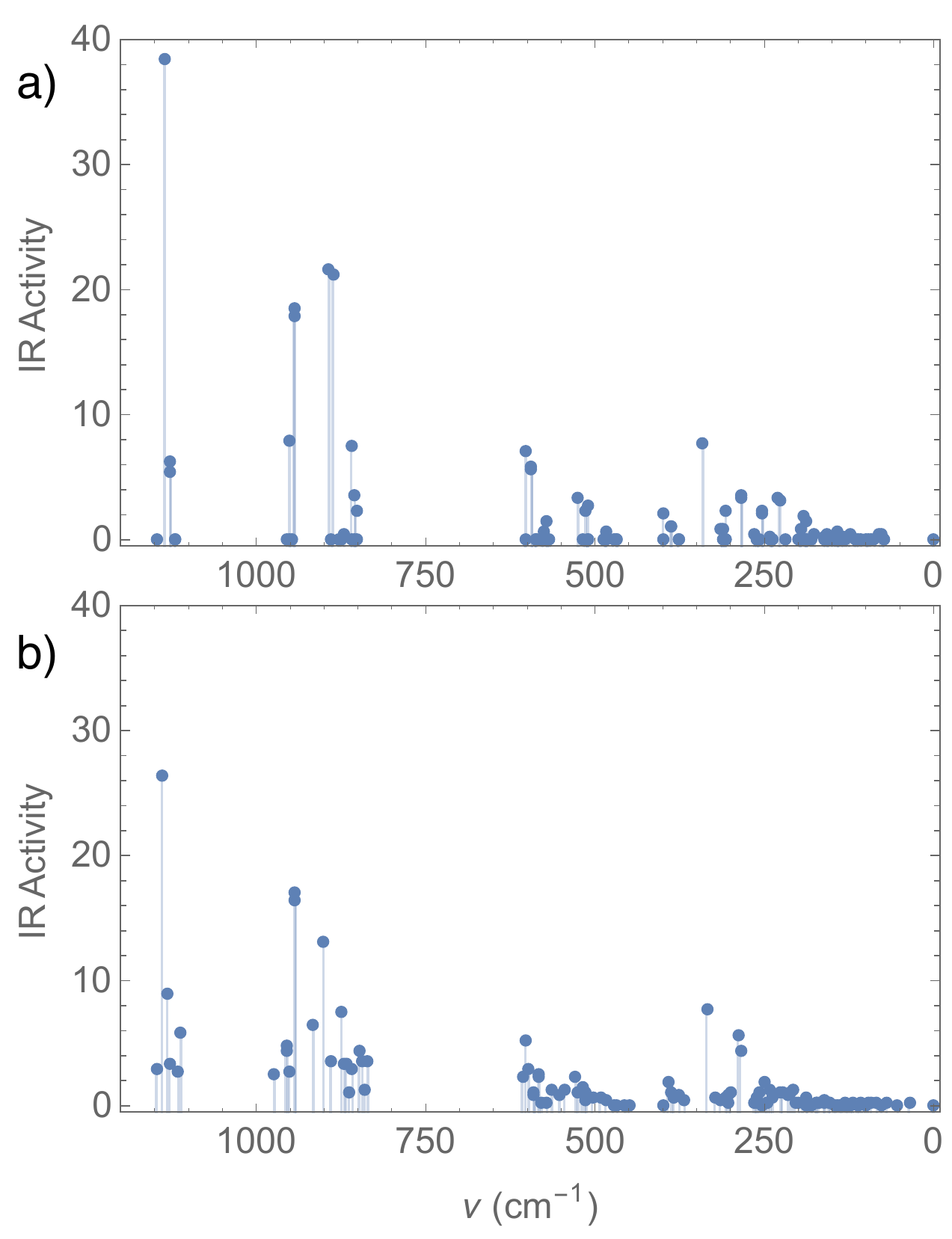}
\caption{Vibrational spectrum of a Posner molecule with a) $\text{S}_6$  and b) $\text{C}_1$ symmetry.  The IR Activity [in (D/\AA)$^2$/amu] of each mode is plotted versus frequency.}
\label{vib}
\end{figure}

The vibrational spectra of both the $\text{S}_6$  structure and a symmetry-broken $\text{C}_1$ structure are shown in Fig.~\ref{vib}.  Note that the $\text{S}_6$  symmetry guarantees that a number of the modes will induce no dipole moment, and thus have zero IR activity.  The corresponding vibrational modes of the $\text{C}_1$ structure are shifted slightly and have a small IR activity, but the overall shape of the spectrum remains the same.  In addition to the plotted modes, we also found modes with small imaginary frequencies.  The $\text{S}_6$ structure has two such modes, and the $\text{C}_1$ has one.  These imaginary frequencies represent soft modes which move between different low-energy states.  Their existence is unsurprising given the presence of many nearly degenerate $\text{C}_1$ structures similar to the $\text{S}_6$ structure.

\subsection{Impurity substitution}
\label{impurities}

During the formation of Posner molecules, ions other than Ca$^{2+}$ and PO$_4^{3-}$ will typically be present in solution, and thus could be substituted for one of the native ions.  We refer to these ionic substitutions as  ``impurities".  Here we consider the energetics of substituting the central calcium ion with either another divalent cation or with two monovalent cations.
In making any comparisons in energies, it is important to take into account the hydration energies of both the
ions to be substituted and of the central calcium ion once removed from the molecule.
In principle the hydration energy of a Posner molecule in solution also needs to be taken into account.
However, it is reasonable to assume that any {\it changes} in this hydration energy upon impurity substitution of the central calcium ion (which is encased inside the molecule) will be small, and hence they are ignored here.
The enthalpy of hydration for a single ion, $\Delta H_\text{hyd}^0$, is defined as the energy change upon taking an ion from a gaseous state to a dilute solution in water.  It is always negative since polar water molecules can considerably reduce their energy by rearranging around the ionic point charge.~\cite{Wulfsberg00}  It is notoriously difficult to calculate hydration enthalpies from first principles~\cite{Soniat15,Gaiduk17}; in this work we take these values from experiment.~\cite{Hydration}

We will denote the enthalpy change due to the substitution of the central calcium ion with either a single divalent cation or of two monovalent cations as $\Delta H_\text{A}$, where A specifies the cation substituted.
$\Delta H_\text{A}$ can be expressed as
\begin{align}
\begin{split}
\Delta H_\text{A}  & =  \Big(\Delta H_f^0[\text{A}_n\text{Ca$_8$(PO$_4$)$_6$}] + \Delta H_\text{hyd}^0 [\text{Ca}^{2+}]\Big) \\
& - \Big(\Delta H_f^0[\text{Ca$_9$(PO$_4$)$_6$}] +  n \Delta H_\text{hyd}^0\big[\text{A}^{(3-n)+}]\Big)
\end{split}
\label{eq:DeltaH}
\end{align}
where $n=1$ corresponds to a single substituted divalent cation, and $n=2$ corresponds to a pair of substituted monovalent cations.
Here $\Delta H_f^0$ is the enthalpy of formation of the Posner molecule with or without the impurity.

Our results for several selected impurities are presented in Table~\ref{impurities_table}.  We find a significant difference in the favorability of various ionic substitutions.  This difference is large enough to outweigh the hydration enthalpy of cations such as Li$^+$ and Fe$^{2+}$, making them highly favored as impurities.  Indeed, the trend is that ions with a stronger tendency to hydrate have an even stronger tendency to substitute for Ca$^{2+}$ in the Posner molecule.  Increased hydration enthalpy is outweighed by increased stability on the central site of the Posner molecule.

These results suggest that if significant concentrations of lithium, iron, or magnesium are present when Posner molecules are formed, they are likely to incorporate into the Posner molecule structure.  This could have a variety of implications.  In the context of calcium phosphate biomineralization, the presence of impurity ions and the nature of their interactions with Posner molecules will have important impacts on Posner-molecule-mediated bone growth.  Additionally, spinful nuclei incorporated as impurities within the Posner molecule will have a significant effect on the phosphorus spin states.

\begin{table}
\begin{tabular*}{0.9\columnwidth}{c @{\extracolsep{\fill}} lrr}
\hline\hline
A$^{2+}$ & $\Delta H_\text{hyd}^0[\text{A$^{2+}$}]$ (eV) & $\Delta H_\text{A$^{2+}$}$ (eV)\\[0.5ex]
\hline
Fe$^{2+}$        & $-$20.21 & $-$1.26 \\
Mg$^{2+}$        & $-$19.96 & $-$1.24 \\
Hg$^{2+}$        & $-$18.96 & $-$0.23 \\
Ca$^{2+}$        & $-$16.37 &  0.00 \\
Pb$^{2+}$        & $-$15.39 &  0.51 \\
\hline\hline
A$^{+}$ & $2 \times \Delta H_\text{hyd}^0[\text{A$^{+}$}]$ (eV) & $\Delta H_\text{A$^{+}$}$ (eV)\\[0.5ex]
\hline
Li$^+$ & $-$10.78 & $-$1.48 \\
Na$^+$ &  $-$8.42 & $-$0.86 \\
K$^+$  &  $-$6.63 &  3.62 \\ [1ex]
\hline
\end{tabular*}
\caption{Energy shift upon substitution of the central Ca$^{2+}$ ion in a Posner molecule with either a divalent cation or with two monovalent cations.  Experimental values for hydration enthalpies ($\Delta H_\text{hyd}^0$) are also listed (Ref.~\onlinecite{Hydration}).
In the case of monovalent cations, the reported value is twice the hydration enthalpy of a single ion.}
\label{impurities_table}
\end{table}

\subsection{Bonding}
\label{bonding}

Aggregation of Posner molecules has been proposed as an intermediate step in biomineralization of amorphous calcium phosphate, a precursor to hydroxyapatite (bone mineral).~\cite{Yin03,Du13,Dey10,Wang12}  We approach this by studying the simplest form of aggregation: pairwise bonding.

\begin{figure}
\includegraphics[width=\columnwidth]{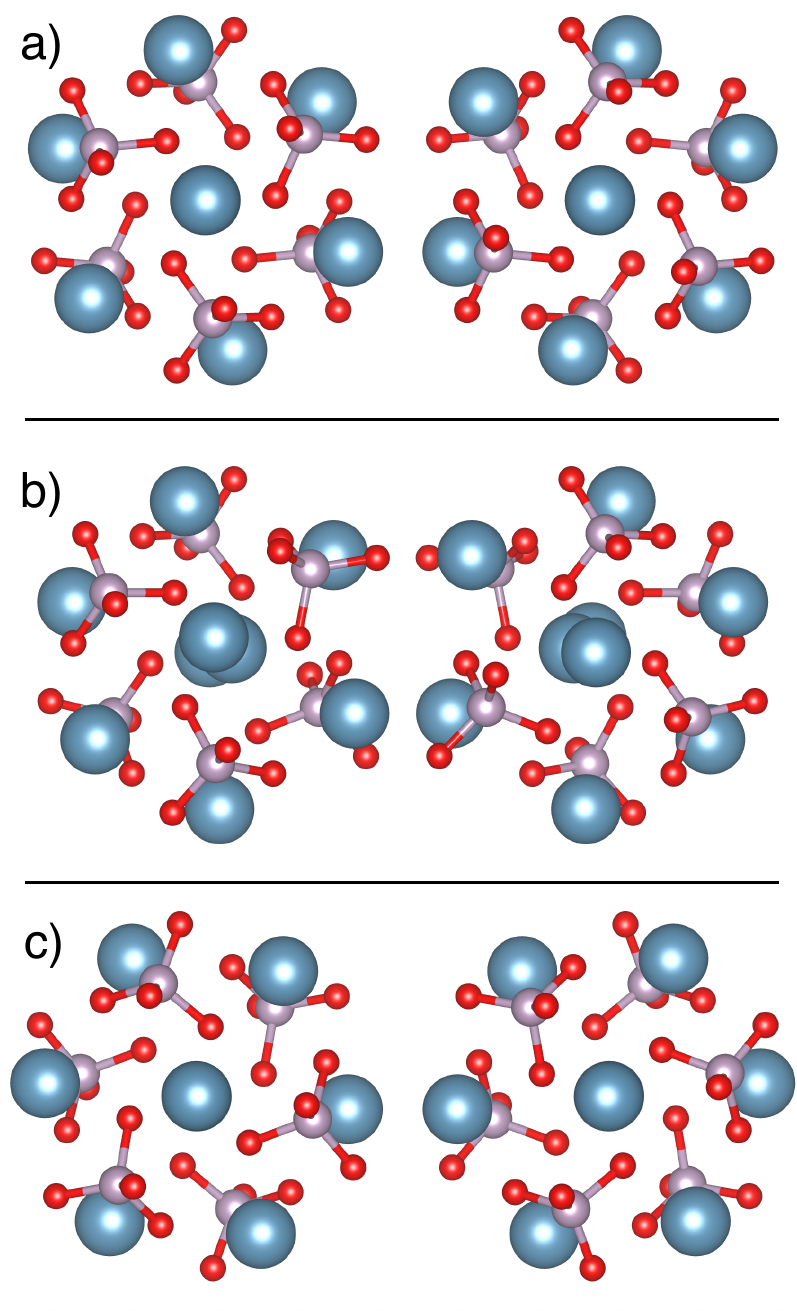}
\caption{Bonding of two Posner molecules.
(a) Bonding configuration for rigid Posner molecules.  The symmetry axes are aligned antiparallel perpendicular to the plane of the page; the two molecules are mirror images of one another.  Where the Posner molecules meet, the Ca$^{2+}$ ions and PO$_4^{3-}$ ions are in different planes perpendicular to the page, keeping a separation between ions of like charge.  This configuration was found through a manual search of bonding distance and relative orientation.  The bonding energy (referenced to two isolated Posner molecules) is 1.04 eV.
(b) Bonded pair of Posner molecules after relaxation starting from the configuration shown in (a).  The bonding energy is 4.99 eV.  The individual molecules are significantly distorted.
(c) Saddle-point configuration in rotation of rigid Posner molecules with respect to one another: $\phi=45^\circ$. }
\label{bond}
\end{figure}

We consider bonding of Posner molecules with the $\text{S}_6$ structure described in Sec.~\ref{Symmetry}.   A variety of bonding orientations for a pair of rigid molecules were tried; the most favorable is depicted in Fig.~\ref{bond}(a) and is similar to the relative ionic positions in hydroxyapatite.
The molecules are mirror images of one another, oriented so that Ca$^{2+}$ ions meet PO$_4^{3-}$ ions, and ions of like charges remain separated.  The distance between centers and the relative orientation of the molecules were varied to find the minimum energy configuration.  The resulting configuration, depicted in Fig.~\ref{bond}(a), has a binding energy of 1.04 eV (referenced to isolated molecules in vacuum).  Starting from this configuration, the constituent ions were allowed to relax.  This relaxation gains another 3.95 eV, for a total bonding energy of 4.99 eV.  The relaxed bonded configuration is shown in Fig.~\ref{bond}(b).

We note that these calculations do not take the presence of solvent into account.  We expect that this is a reasonable approximation when Posner molecules are close enough to bond, since there is not enough space for solvent molecules to enter between the molecules and screen the ionic interaction.  At larger separation, the solvent will likely reduce the bonding tendency, suppressing the long-distance tail of the attraction.

We have also explored the motion of two rigid molecules in a bonded pair with respect to one another.
Specifically, we consider ``rolling without slipping'' rotation, i.e., rotation by both molecules simultaneously in opposite directions, such that the mirror symmetry of the configuration is maintained.  The energy landscape for this rotation is mapped out by repeating the distance-optimization procedure for a set of rotation angles, finding the optimum distance for each orientation.  The saddle-point configuration (shown in Fig.~\ref{bond}(c)) is at a rotation angle of $\phi=45^\circ$, and the rotation barrier is 0.33 eV.
We expect the rotation barrier for rigid molecules is a reasonable approximation in the early stages of the bonding process (before full relaxation).

\subsection{Spin Interactions}
\label{spin_coupling}

\subsubsection{Phosphorus Nuclear Spin Coupling}

Coupling between phosphorus nuclear spins arises due to two factors: magnetic dipole-dipole interaction and ``indirect'' spin-spin coupling.~\cite{NMR}  The rotational motion of the molecule tends to average out the dipole-dipole interaction and the anisotropic part of the indirect coupling, so we only consider the isotropic part of the indirect spin-spin coupling.  This coupling between nuclei $i$ and $j$ is a sum of four terms:
\begin{equation}
J_{ij} = J_{ij}^\text{DSO} + J_{ij}^{PSO} + J_{ij}^{SD} + J_{ij}^{FC} \, ,
\end{equation}
which represent the diamagnetic spin-orbit (DSO), paramagnetic spin-orbit (PSO), spin-dipole (SD), and Fermi contact (FC) terms.~\cite{NMR}
Details of the calculation of nuclear spin-spin couplings were given in Sec.~\ref{methods}.

\begin{figure}
\includegraphics[width=\columnwidth]{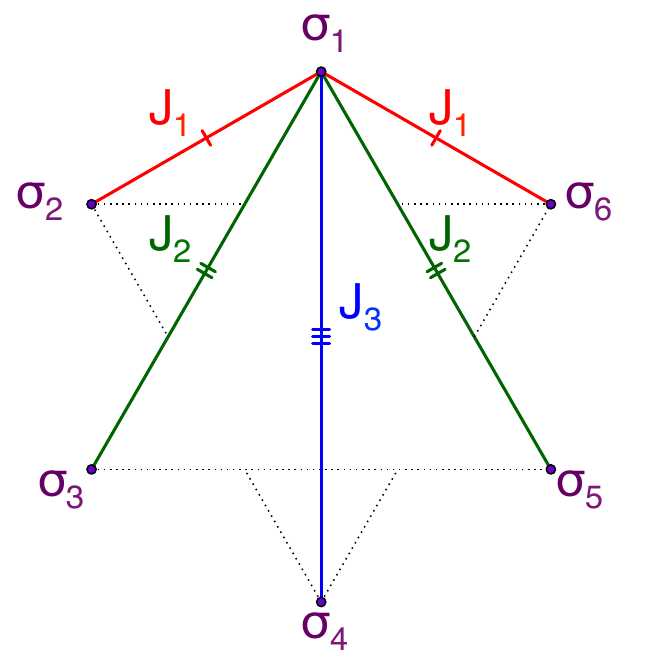}
\caption{Schematic illustration of the configuration of the phosphorus nuclear spins, labeled 1 through 6, with odd spins on the top layer and even on the bottom.  Solid lines indicate the isotropic spin-spin couplings $J_1$, $J_2$, and $J_3$}
\label{illust}
\end{figure}

Adding together these four contributions leads to an effective Heisenberg-like Hamiltonian which describes the interactions between phosphorus nuclear spins:
\begin{equation}
\hat H_0 =  \sum_{i,j} J_{ij}\ \vsigma_i\cdot\vsigma_j \, .
\end{equation}
The nuclear spins of phosphorus are arranged in two equilateral triangles, one on top of the other, centered on the molecule's symmetry axis.
The S$_6$ symmetry restricts the couplings $J_{ij}$ to three values: nearest-neighbor $J_1$, second-nearest-neighbor $J_2$, and third-nearest-neighbor $J_3$, as shown in Fig.~\ref{illust}.  We find $J_1=0.178$ Hz, $J_2=0.145$ Hz, and $J_3 = -0.003$ Hz.

\begin{figure}
\includegraphics[width=\columnwidth]{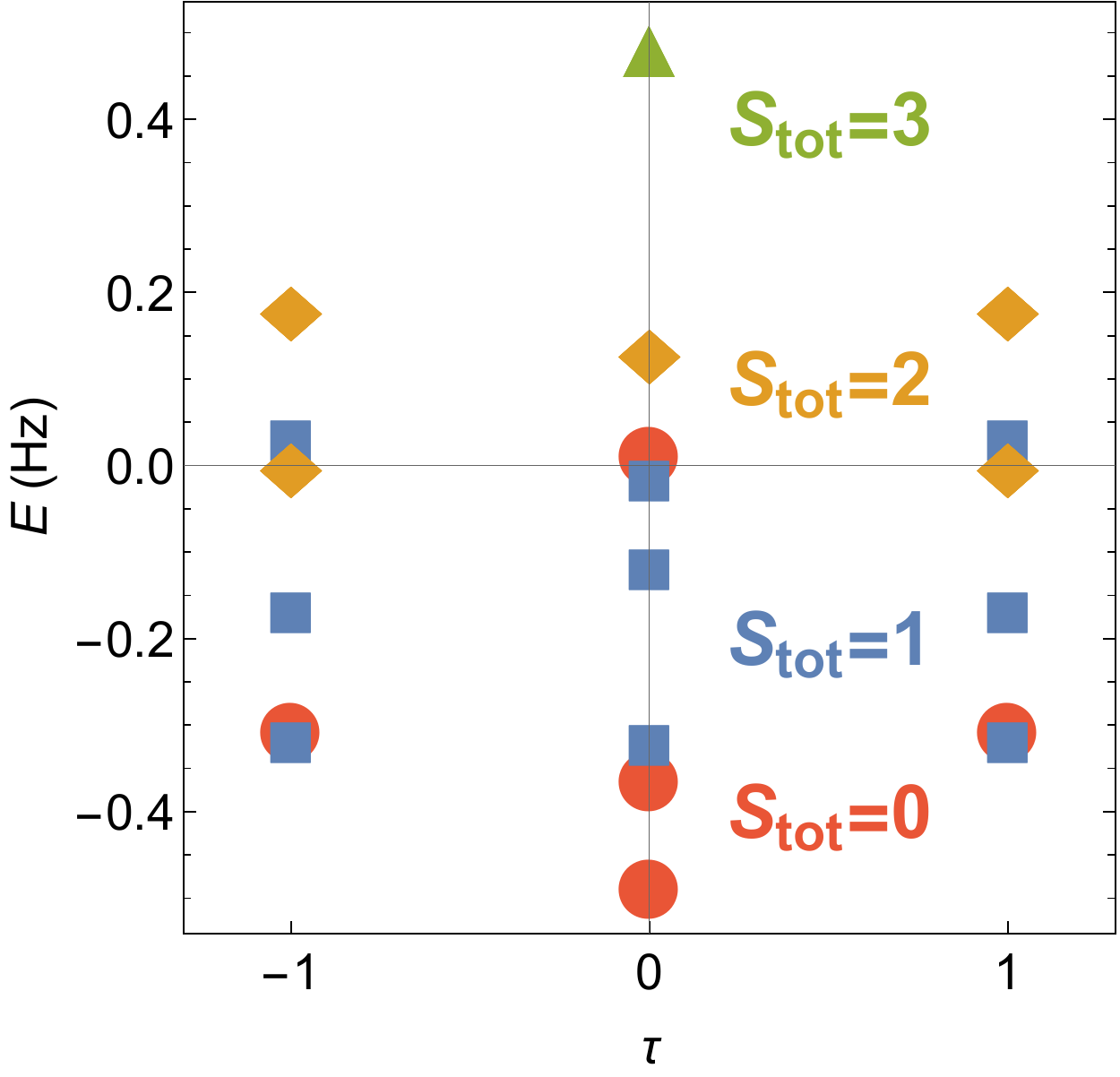}
\caption{Eigenstates of $\hat H$, with energy in Hz on the vertical axis and pseudospin quantum number $\tau$ on the horizontal axis.  The shape and color of the points indicates total spin as shown on the plot.  Each point with spin $S$ has $2S+1$ degeneracy in $S_z$.}
\label{spectrum}
\end{figure}

The threefold rotational symmetry of the Posner molecule ensures that the effective Hamiltonian shares this same symmetry.  Eigenstates of this rotation have an eigenvalue of $e^{i2\pi\tau/3}$, where $\tau$ can take values of $0,\pm1$.  We call this quantum number $\tau$ the ``pseudospin''.
Eigenstates may be expressed using the notation
\begin{equation}\ket\psi = \ket{E,S,S_z,\tau}\end{equation}
where $E$ is the energy (in Hz), $\mathbf{S} = \sum_j \vsigma_j$ is the total spin, and $S_z$ is the $z$ component of total spin.  A plot of the spectrum is shown in Fig.~\ref{spectrum}.

\subsubsection{Pseudospin and Rotations}
\label{Pseudospin}

The transformation properties of the nuclear spin states of a molecule under a symmetry transformation dictate the allowed values of the rotational angular momentum quantum number $L$.
This effect is most dramatic and well-studied in molecular hydrogen ($\text{H}_2$).  Parahydrogen, in which the protons form a spin singlet, is restricted to even values of $L$ by the requirement that the wavefunction be antisymmetric under exchange of the protons.  Orthohydrogen (with a proton spin triplet) is similarly restricted to odd values of $L$.  These spin isomers have different thermodynamic, scattering, and chemical properties.~\cite{Atkins06}

Likewise, for a Posner molecule with threefold rotational symmetry, the Fermi statistics of the $^{31}$P nuclei dictate the allowed rotational angular momentum about the symmetry axis. With three-fold rotational symmetry, the full wavefunction must be unchanged by a rotation by $2\pi/3$, since such a rotation is equivalent to an even number of fermion swapping operations. The pseudospin $\tau$ therefore constrains the angular momentum, $L$, to satisfy: $L + \tau = 0 \text{ mod } 3$.

We propose that this restriction may be important in the case of two Posner molecules ($a,b$) binding together.  Indeed, the recently proposed Quantum Dynamical Selection rule,~\cite{Fisher17} when generalized to the binding of two Posner molecules, predicts that chemical bonding implements a projective measurement onto a state with $\tau_a + \tau_b = 0$ --- essentially a result of the
requirement that binding two Posner molecules stops any relative rotational motion.   If the pair of Posner molecules
subsequently unbinds, this constraint is predicted to be maintained, leaving the two molecules ``pseudospin entangled''.
Thus, pair binding/unbinding of Posner molecules may provide a mechanism
to quantum entangle nuclear spin states in multiple Posner molecules, a necessary precondition for the quantum brain concept.~\cite{Fisher15}

\subsubsection{Decoherence}
\label{Decoherence}

A decoherence time for the spins in a Posner molecule is the NMR spin-lattice relaxation time $T_1$.  The primary mechanism for decoherence is entanglement with external nuclear spins of protons in water molecules external to the Posner molecule.  The $T_1$ due to dipole-dipole interactions between an external spin $M$ (e.g. a proton) and the phosphorus spins (indexed by $I$) is given by the modified Solomon-Bloembergen equation,~\cite{Eldik05}
\begin{align}\label{eq:S-B}
\frac{1}{T_1} = & \sum_{I}\frac{2}{15}M(M+1)C_{DD,I}^2 \left[\frac{\tau_c}{1+(\omega_M - \omega_I)^2\tau_c^2} \right. \\
& \quad \left.+ \frac{3 \tau_c}{1+\omega_I^2\tau_c^2} + \frac{6\tau_c}{1+(\omega_M + \omega_I)^2\tau_c^2}\right] , \nonumber
\end{align}
where $C_{DD,M}$ is the dipole-dipole coupling strength (in Hz) between spin $M$ and $I$, $\tau_c$ is the rotational correlation time, and $\omega_{M,I}$ are the Larmor frequencies of the spins.  We take $M=1/2$, and the correlation time to be given by the thermal rotation frequency $1/\tau_c$
of the Posner molecule.  With a moment of inertia $1.22\times 10^{-43}$ kg m$^2$, $1/\tau_c=2.6 \times 10^{11}$ Hz at 300 K.  This is firmly in the regime $\omega\tau_c \ll 1$.  In this limit, Eq.~(\ref{eq:S-B}) reduces to
\begin{equation}
\frac{1}{T_1} = \sum_I C_{DD,I}^2\tau_c \, .
\end{equation}

As an illustration giving a rough estimate of this coherence time, we consider a proton (perhaps associated with a water molecule in the solvent) as the external spin, located 7\ \AA~from the center of the Posner molecule along the symmetry axis (3.5\ \AA~from the apical Ca$^{2+}$).  This scenario gives
$
T_1 = 1.8\times 10^6 \text{ s} = 21 \text{ days}
$.
Since different pseudospin sectors couple differently to environmental degrees of freedom, the decoherence times for the pseudospin quantum number may be even longer.
The long-lived spin states in the Posner molecule could provide a platform for liquid-state NMR quantum computation, and are also key to the ``quantum brain'' concept set forth in Ref.~\onlinecite{Fisher15}.

\section{Conclusions}

We have explored the structure, symmetry, and spectroscopic fingerprint of the Posner molecule, {\Pos}.  We have shown that Posner molecules are stable in vacuum, and identified S$_6$ symmetry as the prototypical symmetry.  The calculated vibrational spectrum of the Posner molecule may serve as a spectroscopic fingerprint, assisting with the experimental identification of the Posner molecule either \emph{in vitro} or \emph{in vivo}.  Impurity cations can take the place of a central calcium, with implications for both phosphorus spin properties and bone growth.  We find that pairwise Posner molecule bonding is an important process, suggesting avenues for research in bone growth.  Finally, we have shown that the Posner molecule is a promising host for nuclear spins maximally protected from environmental decoherence, with possible implications in liquid-state NMR quantum computation and medical imaging.  We have identified the pseudospin quantum number $\tau$ which could encode long-lived coherent quantum information in the Posner molecule and may provide a mechanism for entangling the molecule's rotational degrees of freedom with its nuclear spin.  This mechanism is central to the Posner molecule's role as a neural qubit in the quantum brain concept.

\acknowledgments

We thank Daniel Ish, Jim Swift, and Leo Radzihovsky for fruitful discussions.
M.P.A.F. is grateful to the Heising-Simons Foundation for support,
to the National Science Foundation for support under Grant No. DMR-14-04230,
and to the Caltech Institute of Quantum Information and Matter, an NSF Physics Frontiers Center with support of the Gordon and Betty Moore Foundation.
Computational resources were provided by the Extreme Science and Engineering Discovery Environment (XSEDE), which is supported by National Science Foundation grant number ACI-1548562, and
by the Center for Scientific Computing from the CNSI, MRL: an NSF MRSEC (DMR-1720256) and NSF CNS-0960316.

\bibliography{Posner_paper}

\begin{thebibliography}{43}%
\makeatletter
\providecommand \@ifxundefined [1]{%
 \@ifx{#1\undefined}
}%
\providecommand \@ifnum [1]{%
 \ifnum #1\expandafter \@firstoftwo
 \else \expandafter \@secondoftwo
 \fi
}%
\providecommand \@ifx [1]{%
 \ifx #1\expandafter \@firstoftwo
 \else \expandafter \@secondoftwo
 \fi
}%
\providecommand \natexlab [1]{#1}%
\providecommand \enquote  [1]{``#1''}%
\providecommand \bibnamefont  [1]{#1}%
\providecommand \bibfnamefont [1]{#1}%
\providecommand \citenamefont [1]{#1}%
\providecommand \href@noop [0]{\@secondoftwo}%
\providecommand \href [0]{\begingroup \@sanitize@url \@href}%
\providecommand \@href[1]{\@@startlink{#1}\@@href}%
\providecommand \@@href[1]{\endgroup#1\@@endlink}%
\providecommand \@sanitize@url [0]{\catcode `\\12\catcode `\$12\catcode
  `\&12\catcode `\#12\catcode `\^12\catcode `\_12\catcode `\%12\relax}%
\providecommand \@@startlink[1]{}%
\providecommand \@@endlink[0]{}%
\providecommand \url  [0]{\begingroup\@sanitize@url \@url }%
\providecommand \@url [1]{\endgroup\@href {#1}{\urlprefix }}%
\providecommand \urlprefix  [0]{URL }%
\providecommand \Eprint [0]{\href }%
\providecommand \doibase [0]{http://dx.doi.org/}%
\providecommand \selectlanguage [0]{\@gobble}%
\providecommand \bibinfo  [0]{\@secondoftwo}%
\providecommand \bibfield  [0]{\@secondoftwo}%
\providecommand \translation [1]{[#1]}%
\providecommand \BibitemOpen [0]{}%
\providecommand \bibitemStop [0]{}%
\providecommand \bibitemNoStop [0]{.\EOS\space}%
\providecommand \EOS [0]{\spacefactor3000\relax}%
\providecommand \BibitemShut  [1]{\csname bibitem#1\endcsname}%
\let\auto@bib@innerbib\@empty
\bibitem [{\citenamefont {Posner}\ and\ \citenamefont
  {Betts}(1975)}]{Posner75}%
  \BibitemOpen
  \bibfield  {author} {\bibinfo {author} {\bibfnamefont {A.~S.}\ \bibnamefont
  {Posner}}\ and\ \bibinfo {author} {\bibfnamefont {F.}~\bibnamefont {Betts}},\
  }\href {\doibase 10.1021/ar50092a003} {\bibfield  {journal} {\bibinfo
  {journal} {Acc. Chem. Res}\ }\textbf {\bibinfo {volume} {8}},\ \bibinfo
  {pages} {273} (\bibinfo {year} {1975})}\BibitemShut {NoStop}%
\bibitem [{\citenamefont {Onuma}\ \emph {et~al.}(1998)\citenamefont {Onuma}, ,\
  and\ \citenamefont {Ito}}]{Onuma98}%
  \BibitemOpen
  \bibfield  {author} {\bibinfo {author} {\bibfnamefont {K.}~\bibnamefont
  {Onuma}}, , \ and\ \bibinfo {author} {\bibfnamefont {A.}~\bibnamefont
  {Ito}},\ }\href {\doibase 10.1021/cm980062c} {\bibfield  {journal} {\bibinfo
  {journal} {Chem. Mater.}\ }\textbf {\bibinfo {volume} {10}},\ \bibinfo
  {pages} {3346} (\bibinfo {year} {1998})}\BibitemShut {NoStop}%
\bibitem [{\citenamefont {Yin}\ and\ \citenamefont {Stott}(2003)}]{Yin03}%
  \BibitemOpen
  \bibfield  {author} {\bibinfo {author} {\bibfnamefont {X.}~\bibnamefont
  {Yin}}\ and\ \bibinfo {author} {\bibfnamefont {M.~J.}\ \bibnamefont
  {Stott}},\ }\href@noop {} {\bibfield  {journal} {\bibinfo  {journal} {J.
  Chem. Phys.}\ }\textbf {\bibinfo {volume} {118}} (\bibinfo {year}
  {2003})}\BibitemShut {NoStop}%
\bibitem [{\citenamefont {Du}\ \emph {et~al.}(2013)\citenamefont {Du},
  \citenamefont {Bian}, \citenamefont {Gou}, \citenamefont {Jiang},
  \citenamefont {Huang}, \citenamefont {Gao}, \citenamefont {Zhao},
  \citenamefont {Wen}, \citenamefont {Zhang},\ and\ \citenamefont
  {Wang}}]{Du13}%
  \BibitemOpen
  \bibfield  {author} {\bibinfo {author} {\bibfnamefont {L.-W.}\ \bibnamefont
  {Du}}, \bibinfo {author} {\bibfnamefont {S.}~\bibnamefont {Bian}}, \bibinfo
  {author} {\bibfnamefont {B.-D.}\ \bibnamefont {Gou}}, \bibinfo {author}
  {\bibfnamefont {Y.}~\bibnamefont {Jiang}}, \bibinfo {author} {\bibfnamefont
  {J.}~\bibnamefont {Huang}}, \bibinfo {author} {\bibfnamefont {Y.-X.}\
  \bibnamefont {Gao}}, \bibinfo {author} {\bibfnamefont {Y.-D.}\ \bibnamefont
  {Zhao}}, \bibinfo {author} {\bibfnamefont {W.}~\bibnamefont {Wen}}, \bibinfo
  {author} {\bibfnamefont {T.-L.}\ \bibnamefont {Zhang}}, \ and\ \bibinfo
  {author} {\bibfnamefont {K.}~\bibnamefont {Wang}},\ }\href {\doibase
  10.1021/cg400498j} {\bibfield  {journal} {\bibinfo  {journal} {Cryst. Growth
  Des}\ }\textbf {\bibinfo {volume} {13}},\ \bibinfo {pages} {3103} (\bibinfo
  {year} {2013})}\BibitemShut {NoStop}%
\bibitem [{\citenamefont {Dey}\ \emph {et~al.}(2010)\citenamefont {Dey},
  \citenamefont {Bomans}, \citenamefont {M{\"u}ller}, \citenamefont {Will},
  \citenamefont {Frederik}, \citenamefont {de~With},\ and\ \citenamefont
  {Sommerdijk}}]{Dey10}%
  \BibitemOpen
  \bibfield  {author} {\bibinfo {author} {\bibfnamefont {A.}~\bibnamefont
  {Dey}}, \bibinfo {author} {\bibfnamefont {P.~H.~H.}\ \bibnamefont {Bomans}},
  \bibinfo {author} {\bibfnamefont {F.~A.}\ \bibnamefont {M{\"u}ller}},
  \bibinfo {author} {\bibfnamefont {J.}~\bibnamefont {Will}}, \bibinfo {author}
  {\bibfnamefont {P.~M.}\ \bibnamefont {Frederik}}, \bibinfo {author}
  {\bibfnamefont {G.}~\bibnamefont {de~With}}, \ and\ \bibinfo {author}
  {\bibfnamefont {N.~A. J.~M.}\ \bibnamefont {Sommerdijk}},\ }\href@noop {}
  {\bibfield  {journal} {\bibinfo  {journal} {Nat. Mater.}\ }\textbf {\bibinfo
  {volume} {9}},\ \bibinfo {pages} {1010} (\bibinfo {year} {2010})}\BibitemShut
  {NoStop}%
\bibitem [{\citenamefont {Wang}\ \emph {et~al.}(2012)\citenamefont {Wang},
  \citenamefont {Li}, \citenamefont {Ruiz-Agudo}, \citenamefont {Putnis},\ and\
  \citenamefont {Putnis}}]{Wang12}%
  \BibitemOpen
  \bibfield  {author} {\bibinfo {author} {\bibfnamefont {L.}~\bibnamefont
  {Wang}}, \bibinfo {author} {\bibfnamefont {S.}~\bibnamefont {Li}}, \bibinfo
  {author} {\bibfnamefont {E.}~\bibnamefont {Ruiz-Agudo}}, \bibinfo {author}
  {\bibfnamefont {C.~V.}\ \bibnamefont {Putnis}}, \ and\ \bibinfo {author}
  {\bibfnamefont {A.}~\bibnamefont {Putnis}},\ }\href {\doibase
  10.1039/C2CE25669J} {\bibfield  {journal} {\bibinfo  {journal} {Cryst. Eng.
  Comm.}\ }\textbf {\bibinfo {volume} {14}},\ \bibinfo {pages} {6252} (\bibinfo
  {year} {2012})}\BibitemShut {NoStop}%
\bibitem [{\citenamefont {Treboux}\ \emph {et~al.}(2000)\citenamefont
  {Treboux}, \citenamefont {Layrolle}, \citenamefont {Kanzaki}, \citenamefont
  {Onuma},\ and\ \citenamefont {Ito}}]{Treboux00}%
  \BibitemOpen
  \bibfield  {author} {\bibinfo {author} {\bibfnamefont {G.}~\bibnamefont
  {Treboux}}, \bibinfo {author} {\bibfnamefont {P.}~\bibnamefont {Layrolle}},
  \bibinfo {author} {\bibfnamefont {N.}~\bibnamefont {Kanzaki}}, \bibinfo
  {author} {\bibfnamefont {K.}~\bibnamefont {Onuma}}, \ and\ \bibinfo {author}
  {\bibfnamefont {A.}~\bibnamefont {Ito}},\ }\href {\doibase 10.1021/jp994399t}
  {\bibfield  {journal} {\bibinfo  {journal} {J. Phys. Chem. A}\ }\textbf
  {\bibinfo {volume} {104}},\ \bibinfo {pages} {5111} (\bibinfo {year}
  {2000})}\BibitemShut {NoStop}%
\bibitem [{\citenamefont {Kanzaki}\ \emph {et~al.}(2001)\citenamefont
  {Kanzaki}, \citenamefont {Treboux}, \citenamefont {Onuma}, \citenamefont
  {Tsutsumi},\ and\ \citenamefont {Ito}}]{Kanzaki01}%
  \BibitemOpen
  \bibfield  {author} {\bibinfo {author} {\bibfnamefont {N.}~\bibnamefont
  {Kanzaki}}, \bibinfo {author} {\bibfnamefont {G.}~\bibnamefont {Treboux}},
  \bibinfo {author} {\bibfnamefont {K.}~\bibnamefont {Onuma}}, \bibinfo
  {author} {\bibfnamefont {S.}~\bibnamefont {Tsutsumi}}, \ and\ \bibinfo
  {author} {\bibfnamefont {A.}~\bibnamefont {Ito}},\ }\href {\doibase
  http://dx.doi.org/10.1016/S0142-9612(01)00039-4} {\bibfield  {journal}
  {\bibinfo  {journal} {Biomaterials}\ }\textbf {\bibinfo {volume} {22}},\
  \bibinfo {pages} {2921 } (\bibinfo {year} {2001})}\BibitemShut {NoStop}%
\bibitem [{\citenamefont {Nicholls}\ and\ \citenamefont
  {Chalmers}(2004)}]{Nicholls04}%
  \BibitemOpen
  \bibfield  {author} {\bibinfo {author} {\bibfnamefont {D.~G.}\ \bibnamefont
  {Nicholls}}\ and\ \bibinfo {author} {\bibfnamefont {S.}~\bibnamefont
  {Chalmers}},\ }\href {\doibase 10.1023/B:JOBB.0000041753.52832.f3} {\bibfield
   {journal} {\bibinfo  {journal} {J. Bioenerg. Biomembr.}\ }\textbf {\bibinfo
  {volume} {36}},\ \bibinfo {pages} {277} (\bibinfo {year} {2004})}\BibitemShut
  {NoStop}%
\bibitem [{Pri()}]{PrivateComm}%
  \BibitemOpen
  \href@noop {} {}\bibinfo {note} {Carol Weingarten and Tobias Fromme, private
  communications (2017)}\BibitemShut {NoStop}%
\bibitem [{\citenamefont {Fisher}(2015)}]{Fisher15}%
  \BibitemOpen
  \bibfield  {author} {\bibinfo {author} {\bibfnamefont {M.~P.~A.}\
  \bibnamefont {Fisher}},\ }\href {\doibase
  http://dx.doi.org/10.1016/j.aop.2015.08.020} {\bibfield  {journal} {\bibinfo
  {journal} {Annals of Physics}\ }\textbf {\bibinfo {volume} {362}},\ \bibinfo
  {pages} {593 } (\bibinfo {year} {2015})}\BibitemShut {NoStop}%
\bibitem [{\citenamefont {Jaszunski}\ \emph {et~al.}(2014)\citenamefont
  {Jaszunski}, \citenamefont {Rizzo},\ and\ \citenamefont {Ruud}}]{NMR}%
  \BibitemOpen
  \bibfield  {author} {\bibinfo {author} {\bibfnamefont {M.}~\bibnamefont
  {Jaszunski}}, \bibinfo {author} {\bibfnamefont {A.}~\bibnamefont {Rizzo}}, \
  and\ \bibinfo {author} {\bibfnamefont {K.}~\bibnamefont {Ruud}},\ }in\ \href
  {\doibase 10.1007/978-94-007-0711-5_11} {\emph {\bibinfo {booktitle}
  {Handbook of Computational Chemistry}}},\ \bibinfo {editor} {edited by\
  \bibinfo {editor} {\bibfnamefont {J.}~\bibnamefont {Leszczynski}}}\ (\bibinfo
   {publisher} {Springer Netherlands},\ \bibinfo {year} {2014})\ pp.\ \bibinfo
  {pages} {361--441}\BibitemShut {NoStop}%
\bibitem [{\citenamefont {Mundy}(1984)}]{Mundy}%
  \BibitemOpen
  \bibfield  {author} {\bibinfo {author} {\bibfnamefont {J.~N.}\ \bibnamefont
  {Mundy}},\ }\href@noop {} {\emph {\bibinfo {title} {Solid State: Nuclear
  Methods}}}\ (\bibinfo  {publisher} {Academic Press},\ \bibinfo {year}
  {1984})\ Chap.\ \bibinfo {chapter} {6.2.1}\BibitemShut {NoStop}%
\bibitem [{\citenamefont {Vandersypen}\ \emph {et~al.}(2001)\citenamefont
  {Vandersypen}, \citenamefont {Steffen}, \citenamefont {Breyta}, \citenamefont
  {Yannoni}, \citenamefont {Sherwood},\ and\ \citenamefont
  {Chuang}}]{Vandersypen01}%
  \BibitemOpen
  \bibfield  {author} {\bibinfo {author} {\bibfnamefont {L.~M.~K.}\
  \bibnamefont {Vandersypen}}, \bibinfo {author} {\bibfnamefont
  {M.}~\bibnamefont {Steffen}}, \bibinfo {author} {\bibfnamefont
  {G.}~\bibnamefont {Breyta}}, \bibinfo {author} {\bibfnamefont {C.~S.}\
  \bibnamefont {Yannoni}}, \bibinfo {author} {\bibfnamefont {M.~H.}\
  \bibnamefont {Sherwood}}, \ and\ \bibinfo {author} {\bibfnamefont {I.~L.}\
  \bibnamefont {Chuang}},\ }\href {\doibase 10.1038/414883a} {\bibfield
  {journal} {\bibinfo  {journal} {Nature}\ }\textbf {\bibinfo {volume} {414}},\
  \bibinfo {pages} {883} (\bibinfo {year} {2001})}\BibitemShut {NoStop}%
\bibitem [{\citenamefont {Zamani}\ \emph {et~al.}(2009)\citenamefont {Zamani},
  \citenamefont {Omrani},\ and\ \citenamefont {Nasab}}]{Zamani09}%
  \BibitemOpen
  \bibfield  {author} {\bibinfo {author} {\bibfnamefont {A.}~\bibnamefont
  {Zamani}}, \bibinfo {author} {\bibfnamefont {G.~R.}\ \bibnamefont {Omrani}},
  \ and\ \bibinfo {author} {\bibfnamefont {M.~M.}\ \bibnamefont {Nasab}},\
  }\href {\doibase http://dx.doi.org/10.1016/j.bone.2008.10.001} {\bibfield
  {journal} {\bibinfo  {journal} {Bone}\ }\textbf {\bibinfo {volume} {44}},\
  \bibinfo {pages} {331 } (\bibinfo {year} {2009})}\BibitemShut {NoStop}%
\bibitem [{\citenamefont {Toxqui}\ and\ \citenamefont
  {Vaquero}(2015)}]{Toxqui15}%
  \BibitemOpen
  \bibfield  {author} {\bibinfo {author} {\bibfnamefont {L.}~\bibnamefont
  {Toxqui}}\ and\ \bibinfo {author} {\bibfnamefont {M.~P.}\ \bibnamefont
  {Vaquero}},\ }\href {\doibase 10.3390/nu7042324} {\bibfield  {journal}
  {\bibinfo  {journal} {Nutrients}\ }\textbf {\bibinfo {volume} {7}},\ \bibinfo
  {pages} {2324} (\bibinfo {year} {2015})}\BibitemShut {NoStop}%
\bibitem [{\citenamefont {Castiglioni}\ \emph {et~al.}(2013)\citenamefont
  {Castiglioni}, \citenamefont {Cazzaniga}, \citenamefont {Albisetti},\ and\
  \citenamefont {Maier}}]{Castiglioni13}%
  \BibitemOpen
  \bibfield  {author} {\bibinfo {author} {\bibfnamefont {S.}~\bibnamefont
  {Castiglioni}}, \bibinfo {author} {\bibfnamefont {A.}~\bibnamefont
  {Cazzaniga}}, \bibinfo {author} {\bibfnamefont {W.}~\bibnamefont
  {Albisetti}}, \ and\ \bibinfo {author} {\bibfnamefont {J.~A.~M.}\
  \bibnamefont {Maier}},\ }\href {\doibase 10.3390/nu5083022} {\bibfield
  {journal} {\bibinfo  {journal} {Nutrients}\ }\textbf {\bibinfo {volume}
  {5}},\ \bibinfo {pages} {3022} (\bibinfo {year} {2013})}\BibitemShut
  {NoStop}%
\bibitem [{\citenamefont {Bl\"ochl}(1994)}]{PAW1}%
  \BibitemOpen
  \bibfield  {author} {\bibinfo {author} {\bibfnamefont {P.~E.}\ \bibnamefont
  {Bl\"ochl}},\ }\href {\doibase 10.1103/PhysRevB.50.17953} {\bibfield
  {journal} {\bibinfo  {journal} {Phys. Rev. B}\ }\textbf {\bibinfo {volume}
  {50}},\ \bibinfo {pages} {17953} (\bibinfo {year} {1994})}\BibitemShut
  {NoStop}%
\bibitem [{\citenamefont {Kresse}\ and\ \citenamefont
  {Furthm\"uller}(1996)}]{VASP4}%
  \BibitemOpen
  \bibfield  {author} {\bibinfo {author} {\bibfnamefont {G.}~\bibnamefont
  {Kresse}}\ and\ \bibinfo {author} {\bibfnamefont {J.}~\bibnamefont
  {Furthm\"uller}},\ }\href {\doibase 10.1103/PhysRevB.54.11169} {\bibfield
  {journal} {\bibinfo  {journal} {Phys. Rev. B}\ }\textbf {\bibinfo {volume}
  {54}},\ \bibinfo {pages} {11169} (\bibinfo {year} {1996})}\BibitemShut
  {NoStop}%
\bibitem [{\citenamefont {Heyd}\ \emph {et~al.}(2003)\citenamefont {Heyd},
  \citenamefont {Scuseria},\ and\ \citenamefont {Ernzerhof}}]{HSE03}%
  \BibitemOpen
  \bibfield  {author} {\bibinfo {author} {\bibfnamefont {J.}~\bibnamefont
  {Heyd}}, \bibinfo {author} {\bibfnamefont {G.~E.}\ \bibnamefont {Scuseria}},
  \ and\ \bibinfo {author} {\bibfnamefont {M.}~\bibnamefont {Ernzerhof}},\
  }\href {\doibase http://dx.doi.org/10.1063/1.1564060} {\bibfield  {journal}
  {\bibinfo  {journal} {J. Chem. Phys.}\ }\textbf {\bibinfo {volume} {118}},\
  \bibinfo {pages} {8207} (\bibinfo {year} {2003})}\BibitemShut {NoStop}%
\bibitem [{\citenamefont {Paier}\ \emph {et~al.}(2006)\citenamefont {Paier},
  \citenamefont {Marsman}, \citenamefont {Hummer}, \citenamefont {Kresse},
  \citenamefont {Gerber},\ and\ \citenamefont {Angyan}}]{HSE06}%
  \BibitemOpen
  \bibfield  {author} {\bibinfo {author} {\bibfnamefont {J.}~\bibnamefont
  {Paier}}, \bibinfo {author} {\bibfnamefont {M.}~\bibnamefont {Marsman}},
  \bibinfo {author} {\bibfnamefont {K.}~\bibnamefont {Hummer}}, \bibinfo
  {author} {\bibfnamefont {G.}~\bibnamefont {Kresse}}, \bibinfo {author}
  {\bibfnamefont {I.~C.}\ \bibnamefont {Gerber}}, \ and\ \bibinfo {author}
  {\bibfnamefont {J.~G.}\ \bibnamefont {Angyan}},\ }\href {\doibase
  http://dx.doi.org/10.1063/1.2403866} {\bibfield  {journal} {\bibinfo
  {journal} {J. Chem. Phys.}\ }\textbf {\bibinfo {volume} {125}},\ \bibinfo
  {eid} {249901} (\bibinfo {year} {2006})}\BibitemShut {NoStop}%
\bibitem [{\citenamefont {Giannozzi}\ \emph {et~al.}(2009)\citenamefont
  {Giannozzi}, \citenamefont {Baroni}, \citenamefont {Bonini}, \citenamefont
  {Calandra}, \citenamefont {Car}, \citenamefont {Cavazzoni}, \citenamefont
  {Ceresoli}, \citenamefont {Chiarotti}, \citenamefont {Cococcioni},
  \citenamefont {Dabo}, \citenamefont {Corso}, \citenamefont {de~Gironcoli},
  \citenamefont {Fabris}, \citenamefont {Fratesi}, \citenamefont {Gebauer},
  \citenamefont {Gerstmann}, \citenamefont {Gougoussis}, \citenamefont
  {Kokalj}, \citenamefont {Lazzeri}, \citenamefont {Martin-Samos},
  \citenamefont {Marzari}, \citenamefont {Mauri}, \citenamefont {Mazzarello},
  \citenamefont {Paolini}, \citenamefont {Pasquarello}, \citenamefont
  {Paulatto}, \citenamefont {Sbraccia}, \citenamefont {Scandolo}, \citenamefont
  {Sclauzero}, \citenamefont {Seitsonen}, \citenamefont {Smogunov},
  \citenamefont {Umari},\ and\ \citenamefont {Wentzcovitch}}]{Giannozzi09}%
  \BibitemOpen
  \bibfield  {author} {\bibinfo {author} {\bibfnamefont {P.}~\bibnamefont
  {Giannozzi}}, \bibinfo {author} {\bibfnamefont {S.}~\bibnamefont {Baroni}},
  \bibinfo {author} {\bibfnamefont {N.}~\bibnamefont {Bonini}}, \bibinfo
  {author} {\bibfnamefont {M.}~\bibnamefont {Calandra}}, \bibinfo {author}
  {\bibfnamefont {R.}~\bibnamefont {Car}}, \bibinfo {author} {\bibfnamefont
  {C.}~\bibnamefont {Cavazzoni}}, \bibinfo {author} {\bibfnamefont
  {D.}~\bibnamefont {Ceresoli}}, \bibinfo {author} {\bibfnamefont {G.~L.}\
  \bibnamefont {Chiarotti}}, \bibinfo {author} {\bibfnamefont {M.}~\bibnamefont
  {Cococcioni}}, \bibinfo {author} {\bibfnamefont {I.}~\bibnamefont {Dabo}},
  \bibinfo {author} {\bibfnamefont {A.~D.}\ \bibnamefont {Corso}}, \bibinfo
  {author} {\bibfnamefont {S.}~\bibnamefont {de~Gironcoli}}, \bibinfo {author}
  {\bibfnamefont {S.}~\bibnamefont {Fabris}}, \bibinfo {author} {\bibfnamefont
  {G.}~\bibnamefont {Fratesi}}, \bibinfo {author} {\bibfnamefont
  {R.}~\bibnamefont {Gebauer}}, \bibinfo {author} {\bibfnamefont
  {U.}~\bibnamefont {Gerstmann}}, \bibinfo {author} {\bibfnamefont
  {C.}~\bibnamefont {Gougoussis}}, \bibinfo {author} {\bibfnamefont
  {A.}~\bibnamefont {Kokalj}}, \bibinfo {author} {\bibfnamefont
  {M.}~\bibnamefont {Lazzeri}}, \bibinfo {author} {\bibfnamefont
  {L.}~\bibnamefont {Martin-Samos}}, \bibinfo {author} {\bibfnamefont
  {N.}~\bibnamefont {Marzari}}, \bibinfo {author} {\bibfnamefont
  {F.}~\bibnamefont {Mauri}}, \bibinfo {author} {\bibfnamefont
  {R.}~\bibnamefont {Mazzarello}}, \bibinfo {author} {\bibfnamefont
  {S.}~\bibnamefont {Paolini}}, \bibinfo {author} {\bibfnamefont
  {A.}~\bibnamefont {Pasquarello}}, \bibinfo {author} {\bibfnamefont
  {L.}~\bibnamefont {Paulatto}}, \bibinfo {author} {\bibfnamefont
  {C.}~\bibnamefont {Sbraccia}}, \bibinfo {author} {\bibfnamefont
  {S.}~\bibnamefont {Scandolo}}, \bibinfo {author} {\bibfnamefont
  {G.}~\bibnamefont {Sclauzero}}, \bibinfo {author} {\bibfnamefont {A.~P.}\
  \bibnamefont {Seitsonen}}, \bibinfo {author} {\bibfnamefont {A.}~\bibnamefont
  {Smogunov}}, \bibinfo {author} {\bibfnamefont {P.}~\bibnamefont {Umari}}, \
  and\ \bibinfo {author} {\bibfnamefont {R.~M.}\ \bibnamefont {Wentzcovitch}},\
  }\href {http://stacks.iop.org/0953-8984/21/i=39/a=395502} {\bibfield
  {journal} {\bibinfo  {journal} {J. Phys. Condens. Matter}\ }\textbf {\bibinfo
  {volume} {21}},\ \bibinfo {pages} {395502} (\bibinfo {year}
  {2009})}\BibitemShut {NoStop}%
\bibitem [{\citenamefont {Vanderbilt}(1990)}]{Vanderbilt90}%
  \BibitemOpen
  \bibfield  {author} {\bibinfo {author} {\bibfnamefont {D.}~\bibnamefont
  {Vanderbilt}},\ }\href {\doibase 10.1103/PhysRevB.41.7892} {\bibfield
  {journal} {\bibinfo  {journal} {Phys. Rev. B}\ }\textbf {\bibinfo {volume}
  {41}},\ \bibinfo {pages} {7892} (\bibinfo {year} {1990})}\BibitemShut
  {NoStop}%
\bibitem [{\citenamefont {Perdew}\ \emph {et~al.}(1996)\citenamefont {Perdew},
  \citenamefont {Burke},\ and\ \citenamefont {Ernzerhof}}]{Perdew96}%
  \BibitemOpen
  \bibfield  {author} {\bibinfo {author} {\bibfnamefont {J.~P.}\ \bibnamefont
  {Perdew}}, \bibinfo {author} {\bibfnamefont {K.}~\bibnamefont {Burke}}, \
  and\ \bibinfo {author} {\bibfnamefont {M.}~\bibnamefont {Ernzerhof}},\ }\href
  {\doibase 10.1103/PhysRevLett.77.3865} {\bibfield  {journal} {\bibinfo
  {journal} {Phys. Rev. Lett.}\ }\textbf {\bibinfo {volume} {77}},\ \bibinfo
  {pages} {3865} (\bibinfo {year} {1996})}\BibitemShut {NoStop}%
\bibitem [{Dal()}]{Dalton1}%
  \BibitemOpen
  \href@noop {} {}\bibinfo {note} {Dalton, a molecular electronic structure
  program, Release Dalton2015.1 (2015), see
  \url{http://daltonprogram.org.}}\BibitemShut {Stop}%
\bibitem [{\citenamefont {Aidas}\ \emph {et~al.}(2015)\citenamefont {Aidas},
  \citenamefont {Angeli}, \citenamefont {Bak}, \citenamefont {Bakken},
  \citenamefont {Bast}, \citenamefont {Boman}, \citenamefont {Christiansen},
  \citenamefont {Cimiraglia}, \citenamefont {Coriani}, \citenamefont {Dahle},
  \citenamefont {Dalskov}, \citenamefont {Ekstr\"{o}m}, \citenamefont
  {Enevoldsen}, \citenamefont {Eriksen}, \citenamefont {Ettenhuber},
  \citenamefont {Fern\'{a}ndez}, \citenamefont {Ferrighi}, \citenamefont
  {Fliegl}, \citenamefont {Frediani}, \citenamefont {Hald}, \citenamefont
  {Halkier}, \citenamefont {H\"{a}ttig}, \citenamefont {Heiberg}, \citenamefont
  {Helgaker}, \citenamefont {Hennum}, \citenamefont {Hettema}, \citenamefont
  {Hjerten\ae{}s}, \citenamefont {H\o{}st}, \citenamefont {H\o{}yvik},
  \citenamefont {Iozzi}, \citenamefont {Jans\'{i}k}, \citenamefont {Jensen},
  \citenamefont {Jonsson}, \citenamefont {J\o{}rgensen}, \citenamefont
  {Kauczor}, \citenamefont {Kirpekar}, \citenamefont {Kj\ae{}rgaard},
  \citenamefont {Klopper}, \citenamefont {Knecht}, \citenamefont {Kobayashi},
  \citenamefont {Koch}, \citenamefont {Kongsted}, \citenamefont {Krapp},
  \citenamefont {Kristensen}, \citenamefont {Ligabue}, \citenamefont
  {Lutn\ae{}s}, \citenamefont {Melo}, \citenamefont {Mikkelsen}, \citenamefont
  {Myhre}, \citenamefont {Neiss}, \citenamefont {Nielsen}, \citenamefont
  {Norman}, \citenamefont {Olsen}, \citenamefont {Olsen}, \citenamefont
  {Osted}, \citenamefont {Packer}, \citenamefont {Pawlowski}, \citenamefont
  {Pedersen}, \citenamefont {Provasi}, \citenamefont {Reine}, \citenamefont
  {Rinkevicius}, \citenamefont {Ruden}, \citenamefont {Ruud}, \citenamefont
  {Rybkin}, \citenamefont {Sa\l{}ek}, \citenamefont {Samson}, \citenamefont
  {de~Mer\'{a}s}, \citenamefont {Saue}, \citenamefont {Sauer}, \citenamefont
  {Schimmelpfennig}, \citenamefont {Sneskov}, \citenamefont {Steindal},
  \citenamefont {Sylvester-Hvid}, \citenamefont {Taylor}, \citenamefont
  {Teale}, \citenamefont {Tellgren}, \citenamefont {Tew}, \citenamefont
  {Thorvaldsen}, \citenamefont {Th\o{}gersen}, \citenamefont {Vahtras},
  \citenamefont {Watson}, \citenamefont {Wilson}, \citenamefont {Ziolkowski},\
  and\ \citenamefont {\AA{}gren}}]{Dalton2}%
  \BibitemOpen
  \bibfield  {author} {\bibinfo {author} {\bibfnamefont {K.}~\bibnamefont
  {Aidas}}, \bibinfo {author} {\bibfnamefont {C.}~\bibnamefont {Angeli}},
  \bibinfo {author} {\bibfnamefont {K.~L.}\ \bibnamefont {Bak}}, \bibinfo
  {author} {\bibfnamefont {V.}~\bibnamefont {Bakken}}, \bibinfo {author}
  {\bibfnamefont {R.}~\bibnamefont {Bast}}, \bibinfo {author} {\bibfnamefont
  {L.}~\bibnamefont {Boman}}, \bibinfo {author} {\bibfnamefont
  {O.}~\bibnamefont {Christiansen}}, \bibinfo {author} {\bibfnamefont
  {R.}~\bibnamefont {Cimiraglia}}, \bibinfo {author} {\bibfnamefont
  {S.}~\bibnamefont {Coriani}}, \bibinfo {author} {\bibfnamefont
  {P.}~\bibnamefont {Dahle}}, \bibinfo {author} {\bibfnamefont {E.~K.}\
  \bibnamefont {Dalskov}}, \bibinfo {author} {\bibfnamefont {U.}~\bibnamefont
  {Ekstr\"{o}m}}, \bibinfo {author} {\bibfnamefont {T.}~\bibnamefont
  {Enevoldsen}}, \bibinfo {author} {\bibfnamefont {J.~J.}\ \bibnamefont
  {Eriksen}}, \bibinfo {author} {\bibfnamefont {P.}~\bibnamefont {Ettenhuber}},
  \bibinfo {author} {\bibfnamefont {B.}~\bibnamefont {Fern\'{a}ndez}}, \bibinfo
  {author} {\bibfnamefont {L.}~\bibnamefont {Ferrighi}}, \bibinfo {author}
  {\bibfnamefont {H.}~\bibnamefont {Fliegl}}, \bibinfo {author} {\bibfnamefont
  {L.}~\bibnamefont {Frediani}}, \bibinfo {author} {\bibfnamefont
  {K.}~\bibnamefont {Hald}}, \bibinfo {author} {\bibfnamefont {A.}~\bibnamefont
  {Halkier}}, \bibinfo {author} {\bibfnamefont {C.}~\bibnamefont {H\"{a}ttig}},
  \bibinfo {author} {\bibfnamefont {H.}~\bibnamefont {Heiberg}}, \bibinfo
  {author} {\bibfnamefont {T.}~\bibnamefont {Helgaker}}, \bibinfo {author}
  {\bibfnamefont {A.~C.}\ \bibnamefont {Hennum}}, \bibinfo {author}
  {\bibfnamefont {H.}~\bibnamefont {Hettema}}, \bibinfo {author} {\bibfnamefont
  {E.}~\bibnamefont {Hjerten\ae{}s}}, \bibinfo {author} {\bibfnamefont
  {S.}~\bibnamefont {H\o{}st}}, \bibinfo {author} {\bibfnamefont {I.-M.}\
  \bibnamefont {H\o{}yvik}}, \bibinfo {author} {\bibfnamefont {M.~F.}\
  \bibnamefont {Iozzi}}, \bibinfo {author} {\bibfnamefont {B.}~\bibnamefont
  {Jans\'{i}k}}, \bibinfo {author} {\bibfnamefont {H.~J.~{\relax Aa}.}\
  \bibnamefont {Jensen}}, \bibinfo {author} {\bibfnamefont {D.}~\bibnamefont
  {Jonsson}}, \bibinfo {author} {\bibfnamefont {P.}~\bibnamefont
  {J\o{}rgensen}}, \bibinfo {author} {\bibfnamefont {J.}~\bibnamefont
  {Kauczor}}, \bibinfo {author} {\bibfnamefont {S.}~\bibnamefont {Kirpekar}},
  \bibinfo {author} {\bibfnamefont {T.}~\bibnamefont {Kj\ae{}rgaard}}, \bibinfo
  {author} {\bibfnamefont {W.}~\bibnamefont {Klopper}}, \bibinfo {author}
  {\bibfnamefont {S.}~\bibnamefont {Knecht}}, \bibinfo {author} {\bibfnamefont
  {R.}~\bibnamefont {Kobayashi}}, \bibinfo {author} {\bibfnamefont
  {H.}~\bibnamefont {Koch}}, \bibinfo {author} {\bibfnamefont {J.}~\bibnamefont
  {Kongsted}}, \bibinfo {author} {\bibfnamefont {A.}~\bibnamefont {Krapp}},
  \bibinfo {author} {\bibfnamefont {K.}~\bibnamefont {Kristensen}}, \bibinfo
  {author} {\bibfnamefont {A.}~\bibnamefont {Ligabue}}, \bibinfo {author}
  {\bibfnamefont {O.~B.}\ \bibnamefont {Lutn\ae{}s}}, \bibinfo {author}
  {\bibfnamefont {J.~I.}\ \bibnamefont {Melo}}, \bibinfo {author}
  {\bibfnamefont {K.~V.}\ \bibnamefont {Mikkelsen}}, \bibinfo {author}
  {\bibfnamefont {R.~H.}\ \bibnamefont {Myhre}}, \bibinfo {author}
  {\bibfnamefont {C.}~\bibnamefont {Neiss}}, \bibinfo {author} {\bibfnamefont
  {C.~B.}\ \bibnamefont {Nielsen}}, \bibinfo {author} {\bibfnamefont
  {P.}~\bibnamefont {Norman}}, \bibinfo {author} {\bibfnamefont
  {J.}~\bibnamefont {Olsen}}, \bibinfo {author} {\bibfnamefont {J.~M.~H.}\
  \bibnamefont {Olsen}}, \bibinfo {author} {\bibfnamefont {A.}~\bibnamefont
  {Osted}}, \bibinfo {author} {\bibfnamefont {M.~J.}\ \bibnamefont {Packer}},
  \bibinfo {author} {\bibfnamefont {F.}~\bibnamefont {Pawlowski}}, \bibinfo
  {author} {\bibfnamefont {T.~B.}\ \bibnamefont {Pedersen}}, \bibinfo {author}
  {\bibfnamefont {P.~F.}\ \bibnamefont {Provasi}}, \bibinfo {author}
  {\bibfnamefont {S.}~\bibnamefont {Reine}}, \bibinfo {author} {\bibfnamefont
  {Z.}~\bibnamefont {Rinkevicius}}, \bibinfo {author} {\bibfnamefont {T.~A.}\
  \bibnamefont {Ruden}}, \bibinfo {author} {\bibfnamefont {K.}~\bibnamefont
  {Ruud}}, \bibinfo {author} {\bibfnamefont {V.~V.}\ \bibnamefont {Rybkin}},
  \bibinfo {author} {\bibfnamefont {P.}~\bibnamefont {Sa\l{}ek}}, \bibinfo
  {author} {\bibfnamefont {C.~C.~M.}\ \bibnamefont {Samson}}, \bibinfo {author}
  {\bibfnamefont {A.~S.}\ \bibnamefont {de~Mer\'{a}s}}, \bibinfo {author}
  {\bibfnamefont {T.}~\bibnamefont {Saue}}, \bibinfo {author} {\bibfnamefont
  {S.~P.~A.}\ \bibnamefont {Sauer}}, \bibinfo {author} {\bibfnamefont
  {B.}~\bibnamefont {Schimmelpfennig}}, \bibinfo {author} {\bibfnamefont
  {K.}~\bibnamefont {Sneskov}}, \bibinfo {author} {\bibfnamefont {A.~H.}\
  \bibnamefont {Steindal}}, \bibinfo {author} {\bibfnamefont {K.~O.}\
  \bibnamefont {Sylvester-Hvid}}, \bibinfo {author} {\bibfnamefont {P.~R.}\
  \bibnamefont {Taylor}}, \bibinfo {author} {\bibfnamefont {A.~M.}\
  \bibnamefont {Teale}}, \bibinfo {author} {\bibfnamefont {E.~I.}\ \bibnamefont
  {Tellgren}}, \bibinfo {author} {\bibfnamefont {D.~P.}\ \bibnamefont {Tew}},
  \bibinfo {author} {\bibfnamefont {A.~J.}\ \bibnamefont {Thorvaldsen}},
  \bibinfo {author} {\bibfnamefont {L.}~\bibnamefont {Th\o{}gersen}}, \bibinfo
  {author} {\bibfnamefont {O.}~\bibnamefont {Vahtras}}, \bibinfo {author}
  {\bibfnamefont {M.~A.}\ \bibnamefont {Watson}}, \bibinfo {author}
  {\bibfnamefont {D.~J.~D.}\ \bibnamefont {Wilson}}, \bibinfo {author}
  {\bibfnamefont {M.}~\bibnamefont {Ziolkowski}}, \ and\ \bibinfo {author}
  {\bibfnamefont {H.}~\bibnamefont {\AA{}gren}},\ }\href {\doibase
  10.1002/wcms.1172} {\bibfield  {journal} {\bibinfo  {journal} {WIREs
  Comput.~Mol.~Sci.}\ }\textbf {\bibinfo {volume} {4}},\ \bibinfo {pages} {269}
  (\bibinfo {year} {2015})}\BibitemShut {NoStop}%
\bibitem [{\citenamefont {Becke}(1993)}]{Becke93}%
  \BibitemOpen
  \bibfield  {author} {\bibinfo {author} {\bibfnamefont {A.~D.}\ \bibnamefont
  {Becke}},\ }\href@noop {} {\bibfield  {journal} {\bibinfo  {journal} {J.
  Chem. Phys.}\ }\textbf {\bibinfo {volume} {98}} (\bibinfo {year}
  {1993})}\BibitemShut {NoStop}%
\bibitem [{\citenamefont {Helgaker}\ \emph {et~al.}(2000)\citenamefont
  {Helgaker}, \citenamefont {Watson},\ and\ \citenamefont
  {Handy}}]{Helgaker00}%
  \BibitemOpen
  \bibfield  {author} {\bibinfo {author} {\bibfnamefont {T.}~\bibnamefont
  {Helgaker}}, \bibinfo {author} {\bibfnamefont {M.}~\bibnamefont {Watson}}, \
  and\ \bibinfo {author} {\bibfnamefont {N.~C.}\ \bibnamefont {Handy}},\ }\href
  {\doibase 10.1063/1.1321296} {\bibfield  {journal} {\bibinfo  {journal} {J.
  Chem. Phys.}\ }\textbf {\bibinfo {volume} {113}},\ \bibinfo {pages} {9402}
  (\bibinfo {year} {2000})}\BibitemShut {NoStop}%
\bibitem [{\citenamefont {Helgaker}\ \emph {et~al.}(2008)\citenamefont
  {Helgaker}, \citenamefont {Jaszuński},\ and\ \citenamefont
  {Pecul}}]{Helgaker08}%
  \BibitemOpen
  \bibfield  {author} {\bibinfo {author} {\bibfnamefont {T.}~\bibnamefont
  {Helgaker}}, \bibinfo {author} {\bibfnamefont {M.}~\bibnamefont
  {Jaszuński}}, \ and\ \bibinfo {author} {\bibfnamefont {M.}~\bibnamefont
  {Pecul}},\ }\href {\doibase 10.1016/j.pnmrs.2008.02.002} {\bibfield
  {journal} {\bibinfo  {journal} {Prog. Nucl. Mag. Res. Sp.}\ }\textbf
  {\bibinfo {volume} {53}},\ \bibinfo {pages} {249 } (\bibinfo {year}
  {2008})}\BibitemShut {NoStop}%
\bibitem [{\citenamefont {Hariharan}\ and\ \citenamefont
  {Pople}(1973)}]{Pople73}%
  \BibitemOpen
  \bibfield  {author} {\bibinfo {author} {\bibfnamefont {P.}~\bibnamefont
  {Hariharan}}\ and\ \bibinfo {author} {\bibfnamefont {J.}~\bibnamefont
  {Pople}},\ }\href {\doibase 10.1007/BF00533485} {\bibfield  {journal}
  {\bibinfo  {journal} {Theor. chim. acta}\ }\textbf {\bibinfo {volume} {28}},\
  \bibinfo {pages} {213} (\bibinfo {year} {1973})}\BibitemShut {NoStop}%
\bibitem [{\citenamefont {Rassolov}\ \emph {et~al.}(1998)\citenamefont
  {Rassolov}, \citenamefont {Pople}, \citenamefont {Ratner},\ and\
  \citenamefont {Windus}}]{Pople98}%
  \BibitemOpen
  \bibfield  {author} {\bibinfo {author} {\bibfnamefont {V.~A.}\ \bibnamefont
  {Rassolov}}, \bibinfo {author} {\bibfnamefont {J.~A.}\ \bibnamefont {Pople}},
  \bibinfo {author} {\bibfnamefont {M.~A.}\ \bibnamefont {Ratner}}, \ and\
  \bibinfo {author} {\bibfnamefont {T.~L.}\ \bibnamefont {Windus}},\
  }\href@noop {} {\bibfield  {journal} {\bibinfo  {journal} {J. Chem. Phys.}\
  }\textbf {\bibinfo {volume} {109}} (\bibinfo {year} {1998})}\BibitemShut
  {NoStop}%
\bibitem [{\citenamefont {Jensen}(2006)}]{Jensen06}%
  \BibitemOpen
  \bibfield  {author} {\bibinfo {author} {\bibfnamefont {F.}~\bibnamefont
  {Jensen}},\ }\href {\doibase 10.1021/ct600166u} {\bibfield  {journal}
  {\bibinfo  {journal} {J. Chem. Theory Comput.}\ }\textbf {\bibinfo {volume}
  {2}},\ \bibinfo {pages} {1360} (\bibinfo {year} {2006})}\BibitemShut
  {NoStop}%
\bibitem [{\citenamefont {Momma}\ and\ \citenamefont {Izumi}(2011)}]{VESTA}%
  \BibitemOpen
  \bibfield  {author} {\bibinfo {author} {\bibfnamefont {K.}~\bibnamefont
  {Momma}}\ and\ \bibinfo {author} {\bibfnamefont {F.}~\bibnamefont {Izumi}},\
  }\href {\doibase 10.1107/S0021889811038970} {\bibfield  {journal} {\bibinfo
  {journal} {J. Appl. Crystallogr.}\ }\textbf {\bibinfo {volume} {44}},\
  \bibinfo {pages} {1272} (\bibinfo {year} {2011})}\BibitemShut {NoStop}%
\bibitem [{\citenamefont {Grases}\ \emph {et~al.}(2014)\citenamefont {Grases},
  \citenamefont {Zelenková},\ and\ \citenamefont {Söhnel}}]{Grases14}%
  \BibitemOpen
  \bibfield  {author} {\bibinfo {author} {\bibfnamefont {F.}~\bibnamefont
  {Grases}}, \bibinfo {author} {\bibfnamefont {M.}~\bibnamefont {Zelenková}},
  \ and\ \bibinfo {author} {\bibfnamefont {O.}~\bibnamefont {Söhnel}},\ }\href
  {\doibase 10.1007/s00240-013-0611-6} {\bibfield  {journal} {\bibinfo
  {journal} {Urolithiasis}\ }\textbf {\bibinfo {volume} {42}},\ \bibinfo
  {pages} {9} (\bibinfo {year} {2014})}\BibitemShut {NoStop}%
\bibitem [{\citenamefont {Wilson}(1955)}]{Wilson}%
  \BibitemOpen
  \bibfield  {author} {\bibinfo {author} {\bibfnamefont {E.~B.}\ \bibnamefont
  {Wilson}},\ }\href@noop {} {\emph {\bibinfo {title} {Molecular vibrations;
  the theory of infrared and Raman vibrational spectra}}}\ (\bibinfo
  {publisher} {McGraw-Hill},\ \bibinfo {year} {1955})\BibitemShut {NoStop}%
\bibitem [{\citenamefont {Porezag}\ and\ \citenamefont
  {Pederson}(1996)}]{Porezag96}%
  \BibitemOpen
  \bibfield  {author} {\bibinfo {author} {\bibfnamefont {D.}~\bibnamefont
  {Porezag}}\ and\ \bibinfo {author} {\bibfnamefont {M.~R.}\ \bibnamefont
  {Pederson}},\ }\href {\doibase 10.1103/PhysRevB.54.7830} {\bibfield
  {journal} {\bibinfo  {journal} {Phys. Rev. B}\ }\textbf {\bibinfo {volume}
  {54}},\ \bibinfo {pages} {7830} (\bibinfo {year} {1996})}\BibitemShut
  {NoStop}%
\bibitem [{\citenamefont {Wulfsberg}(2000)}]{Wulfsberg00}%
  \BibitemOpen
  \bibfield  {author} {\bibinfo {author} {\bibfnamefont {G.}~\bibnamefont
  {Wulfsberg}},\ }\href@noop {} {\emph {\bibinfo {title} {Inorganic
  Chemistry}}}\ (\bibinfo  {publisher} {University Science Books},\ \bibinfo
  {year} {2000})\BibitemShut {NoStop}%
\bibitem [{\citenamefont {Soniat}\ \emph {et~al.}(2015)\citenamefont {Soniat},
  \citenamefont {Rogers},\ and\ \citenamefont {Rempe}}]{Soniat15}%
  \BibitemOpen
  \bibfield  {author} {\bibinfo {author} {\bibfnamefont {M.}~\bibnamefont
  {Soniat}}, \bibinfo {author} {\bibfnamefont {D.~M.}\ \bibnamefont {Rogers}},
  \ and\ \bibinfo {author} {\bibfnamefont {S.~B.}\ \bibnamefont {Rempe}},\
  }\href {\doibase 10.1021/acs.jctc.5b00357} {\bibfield  {journal} {\bibinfo
  {journal} {J. Chem. Theory Comput.}\ }\textbf {\bibinfo {volume} {11}},\
  \bibinfo {pages} {2958} (\bibinfo {year} {2015})}\BibitemShut {NoStop}%
\bibitem [{\citenamefont {Gaiduk}\ and\ \citenamefont
  {Galli}(2017)}]{Gaiduk17}%
  \BibitemOpen
  \bibfield  {author} {\bibinfo {author} {\bibfnamefont {A.~P.}\ \bibnamefont
  {Gaiduk}}\ and\ \bibinfo {author} {\bibfnamefont {G.}~\bibnamefont {Galli}},\
  }\href {\doibase 10.1021/acs.jpclett.7b00239} {\bibfield  {journal} {\bibinfo
   {journal} {J. Phys. Chem. Lett.}\ }\textbf {\bibinfo {volume} {8}},\
  \bibinfo {pages} {1496} (\bibinfo {year} {2017})}\BibitemShut {NoStop}%
\bibitem [{\citenamefont {Smith}(1977)}]{Hydration}%
  \BibitemOpen
  \bibfield  {author} {\bibinfo {author} {\bibfnamefont {D.~W.}\ \bibnamefont
  {Smith}},\ }\href {\doibase 10.1021/ed054p540} {\bibfield  {journal}
  {\bibinfo  {journal} {J. Chem. Educ.}\ }\textbf {\bibinfo {volume} {54}},\
  \bibinfo {pages} {540} (\bibinfo {year} {1977})}\BibitemShut {NoStop}%
\bibitem [{\citenamefont {Atkins}\ and\ \citenamefont
  {de~Paula}(2006)}]{Atkins06}%
  \BibitemOpen
  \bibfield  {author} {\bibinfo {author} {\bibfnamefont {P.}~\bibnamefont
  {Atkins}}\ and\ \bibinfo {author} {\bibfnamefont {J.}~\bibnamefont
  {de~Paula}},\ }\href@noop {} {\emph {\bibinfo {title} {Physical Chemistry,
  8th Edition}}}\ (\bibinfo  {publisher} {W.H.Freeman},\ \bibinfo {year}
  {2006})\BibitemShut {NoStop}%
\bibitem [{\citenamefont {Fisher}\ and\ \citenamefont
  {Radzihovsky}(2017)}]{Fisher17}%
  \BibitemOpen
  \bibfield  {author} {\bibinfo {author} {\bibfnamefont {M.~P.~A.}\
  \bibnamefont {Fisher}}\ and\ \bibinfo {author} {\bibfnamefont
  {L.}~\bibnamefont {Radzihovsky}},\ }\href {https://arxiv.org/abs/1707.05320}
  {\bibfield  {journal} {\bibinfo  {journal} {arXiv:1707.05320
  [physics.chem-ph]}\ } (\bibinfo {year} {2017})}\BibitemShut {NoStop}%
\bibitem [{\citenamefont {van Eldik}(2005)}]{Eldik05}%
  \BibitemOpen
  \bibfield  {author} {\bibinfo {author} {\bibfnamefont {R.}~\bibnamefont {van
  Eldik}},\ }\href@noop {} {\emph {\bibinfo {title} {Advances in Inorganic
  Chemistry, Vol. 57}}}\ (\bibinfo  {publisher} {Academic Press},\ \bibinfo
  {year} {2005})\BibitemShut {NoStop}%
\end{thebibliography}%

\end{document}